\documentclass[aip,cha,floatfix,reprint]{revtex4-1}
\usepackage{color}
\usepackage{graphicx}

\newcommand{\dt}{T}

\begin{document}

\title{Defining Chaos} 

\author{Brian R. Hunt}
\affiliation{University of Maryland, College Park MD 20742, USA}

\author{Edward Ott}
\affiliation{University of Maryland, College Park MD 20742, USA}

\date{10 April 2015}

\begin{abstract}
In this paper we propose, discuss and illustrate a computationally feasible definition of chaos which can be applied very generally to situations that are commonly encountered, including attractors, repellers and non-periodically forced systems.  This definition is based on an entropy-like quantity, which we call ``expansion entropy'', and we define chaos as occurring when this quantity is positive.  We relate and compare expansion entropy to the well-known concept of topological entropy, to which it is equivalent under appropriate conditions.  We also present example illustrations, discuss computational implementations, and point out issues arising from attempts at giving definitions of chaos that are not entropy-based.
\end{abstract}


\maketitle 

\begin{quotation}
Toward the end of the 19th century, Poincar\'e demonstrated the occurrence of extremely complicated orbits in the Newtonian dynamics of three gravitationally attracting bodies.  This complexity is now called chaos and has received a vast amount of attention since Poincar\'e's early discovery.  In spite of this abundant past and current work, there is still no broadly applicable, convenient, generally accepted definition of the term chaos.  In this paper, we advocate a particular entropy-based definition that appears to be very simple, while, at the same time, is readily accessible to numerical computation, and can be very generally applied to a variety of often-encountered situations, including attractors, repellers, and non-periodically forced systems.  We also review and compare various previous definitions of chaos.
\end{quotation}


\section{Initial Discussion}
\label{sec1}

While the word chaos is widely used in science and mathematics, there are a variety of ways of defining it.  Thus, for this 25th anniversary issue of the journal CHAOS, we are motivated to review issues that arise when attempting to formulate a generally applicable definition of chaos, and to advocate a particular entropy-based definition that seems to us to be especially apt.  We also relate our proposed definition to previous definitions.

Intuitively, perhaps the two most prominent (not necessarily independent) attributes of what scientists commonly think of as chaos are the presence of complex orbit structure and extreme sensitivity of orbits to small perturbations.  Indeed, in the paper by Li and Yorke\cite{LY} where the term chaos was introduced in its now widely accepted nonlinear dynamics context, the term was motivated by the simultaneous presence of unstable periodic orbits of all periods, as well as an uncountable infinity of non-periodic orbits.  Thus, Li and Yorke's introduction of this terminology was motivated by the chaos attribute of complex orbit structure.  On the other hand, Lorenz\cite{L} was concerned with weather forecasting and accordingly focused on the chaos attribute of temporally exponential increase of the sensitivity of orbit locations to small initial perturbations.  As we will discuss, these two attributes can be viewed as ``two sides of the same coin''.

We think of a definition of chaos as being ``good'' if it conforms to common intuitive notions of chaos (such as complex orbit structure and orbit sensitivity) and, at the same time, has the following three desirable features:

\begin{itemize}
\item	\emph{Generality}:  The definition should work for almost all the examples that typical readers of this journal are likely to judge as chaotic.
\item	\emph{Simplicity}: The definition should be fairly concise and not too technical.
\item	\emph{Computability}: The definition should allow a practical, straightforward computational implementation for discerning the existence of chaos in a model.
\end{itemize}

Considering the issue of generality, one would like a definition of chaos to be applicable not only to attractors, but also to non-attracting sets, often called repellers.  With respect to chaotic repellers\cite{GOY,KG,LT}, we note that they are central to the physically relevant topics of fractal basin boundaries\cite{MGOY}, chaotic transients, and chaotic scattering\cite{TO}, occurring, for example in fluid dynamics\cite{SYE,SKG}, celestial mechanics\cite{JRLMFU}, chemistry\cite{GE}, and atomic physics\cite{DD}.

Furthermore, again considering the issue of generality, due to their common occurrence in applications, we desire that our definition of chaos be applicable to non-autonomous dynamical systems (i.e., systems that are externally forced by time dependent inputs), including external inputs that are temporally quasi-periodic\cite{FKP}, stochastic\cite{LY1,LY2,YOC}, or are themselves chaotic.  Here physical examples include quasi-periodic forcing of atmospheric jets\cite{RBKOU}, quasi-periodic forcing of stellar luminosity variation by two superposed stellar modal oscillations\cite{LKKHLD,Mo}, advective transport in fluids with temporally and spatially irregular flow fields\cite{VAO,SO,HY,VHG,MHPRS}, and phase synchronism of chaos by noisy or chaotic drives\cite{PROK}.  We emphasize that, when considering externally forced systems, we are interested in identifying chaos in the response of the system to a particular realization of the forcing, not in characterizing whether the forcing is chaotic.

An important point for consideration of non-periodically forced chaotic systems is that the notion of a compact invariant set, which is typically used in definitions of chaos for autonomous systems (including Poincar\'e maps of periodically forced systems), may not be appropriate or convenient for situations with non-periodic forcing.  Furthermore, in practice, it may be difficult to locate or detect an invariant set that is not an attractor.  Thus, rather than defining chaos for an invariant set, we will instead consider a notion of chaos for the dynamics within any given bounded positive-volume subset $S$ of the state space.  We call such a set $S$ a \emph{restraining region}.  For autonomous systems, chaos for an invariant set can be detected by taking $S$ to be a neighborhood of the desired invariant set.

In our opinion, the currently most satisfactory way of defining chaos for autonomous systems is by the existence of positive topological entropy or metric entropy.  We note, however, that the standard definitions of these entropies are quite difficult to straightforwardly implement in a numerical procedure.  In addition, while generalizations to the original definitions of topological and metric entropy have been proposed, we view it as desirable to have a relatively simple definition that is applicable very broadly.  However, we do not address here the question of identifying chaos in experimental data, which presents additional challenges, especially in the cases of non-attracting sets and externally forced systems.

Motivated by the considerations above, in Sec.~\ref{sec2} we introduce and discuss the definition of an alternate entropy quantity that we call ``expansion entropy''.  The expansion entropy of an $n$-dimensional dynamical system on a restraining region $S$ is the difference between two asymptotic exponential rates: first, the maximum over $d \leq n$ of the rate at which the system expands $d$-dimensional volume within $S$; and second, the rate at which $n$-dimensional volume leaves $S$ (this rate is $0$ for an invariant set).  \emph{We define chaos as the existence of positive expansion entropy on a given restraining region.}  Expansion entropy generalizes (to nonautonomous systems and noninvariant restraining regions) a quantity that was formulated by Sacksteder and Shub\cite{SS} in the case of an autonomous system on a compact manifold.  In this restricted case, by the results of Kozlovski\cite{K}, expansion entropy is equal to topological entropy for infinitely differentiable maps.  In Sec.~\ref{sec2.6} we present examples of the application of our definition of expansion entropy to various systems, and also provide illustrative numerical evaluations of expansion entropy for some of these examples.  Section~\ref{sec3} discusses topological entropy and previous work on computation of this quantity.  Section~\ref{sec4} discusses issues that arise in previous non-entropy-based definitions of chaos.

\section{Expansion Entropy}
\label{sec2}

\subsection{Definition}
\label{sec2.1}

Our definition of expansion entropy, which we denote $H_0$, is closely related to previous definitions of topological entropy, to which it is equivalent under appropriate conditions (see Sec.~\ref{sec3}).  Expansion entropy uses the linearization of the dynamical system and a notion of volume on its state space; thus, unlike topological entropy, it is defined only for smooth dynamical systems.  On the other hand, expansion entropy does not require the identification of a compact invariant set.  As we will discuss, the differences in the definitions may make the criterion $H_0 > 0$ attractive as a general definition of chaos in smooth dynamical systems.

We assume that the state space of the dynamical system is a finite-dimensional manifold $M$.  (For example, if the manifold $M$ is $n$-dimensional Euclidean space, the state $x$ at a given time is a vector $[x_1,x_2,\ldots,x_n]$ where each $x_i$ is a real number.  If some of the coordinates are angle variables, they can be taken modulo $2\pi$, resulting in manifolds such as a circle, cylinder or torus.)  We write $\mu(S)$ for the volume\cite{footnote1} of a subset $S$ of $M$, and write $d\mu(x)$ for integration with respect to this volume; for $n$-dimensional Euclidean space, one can take $d\mu(x) = d^n x$.  Given a dynamical system on $M$, we will use an integral to define its expansion entropy on a closed subset $S$ (the restraining region) that has positive, finite volume.  The set $S$ need not be invariant under the system.

We consider a deterministic dynamical system to be defined by an evolution operator, by which we mean a family $f$ of maps $f_{t',t} : M \to M$, with the interpretation that if $x$ and $x'$ are the states of the system at times $t$ and $t'$, respectively, then $x' = f_{t',t}(x)$.  (For example, $f_{t',t}$ could represent the solution from time $t$ to $t'$ of a system of differential equations.)  The family $f$ must satisfy the identities $f_{t,t}(x) = x$ and $f_{t'',t}(x) = f_{t'',t'}(f_{t',t}(x))$.  The maps $f_{t',t}$ are defined for $t',t$ being integer-valued (discrete time) or being real numbers (continuous time), with the restriction that $t' \geq t$ if the system is noninvertible.  We allow the system to be nonautonomous, including the case where $f_{t',t}$ is a realization of a stochastic dynamical system.\cite{footnote2}  If the system is autonomous, then $f_{t',t}$ depends only on $t' - t$, and in this case we will often write $f_{t',t} = f_\dt$ where $\dt = t' - t$.  Regardless, we assume that $f_{t',t}$ is a differentiable function of $x$.

Recall that the singular values of a matrix $A$ are the square roots of the eigenvalues of $A^\top A$.  Thinking of $A$ as a linear transformation, the image of the unit ball under $A$ is an ellipsoid, and the singular values of $A$ are the semiaxes of this ellipsoid.  Let $G(A)$ be the product of the singular values of $A$ that are greater than $1$; if none of the singular values are greater than $1$, let $G(A) = 1$.  Then $G(A)$ is roughly the number of $\epsilon$-balls needed to cover the image of an $\epsilon$-ball under $A$.

If the matrix $A$ is $n \times n$, consider a $d$-dimensional plane $P_d$ in $n$-dimensional Euclidean space, where $d \leq n$.  Let $W$ be a $d$-dimensional ball in $P_d$, let $A(W)$ denote the image of $W$ under $A$, and let $\mu_d$ denote $d$-dimensional volume.  Then $G(A)$ is the maximum over orientations of $P_d$ and over $d$ of $\mu_d(A(W))/\mu_d(W)$.  Thus, \emph{$G(A)$ is the largest possible growth ratio of $d$-dimensional volumes under $A$}.  Below we will apply $G$ to the derivative matrix $Df_{t',t}$, in which case it represents a \emph{local} volume growth ratio for the (typically nonlinear) map $f_{t',t}$.

Let $S_{t',t}$ be the set of $x$ such that $f_{t'',t}(x) \in S$ for all $t''$ between $t$ and $t'$ (that is, the trajectory of $x$ from $t$ to $t'$ under $f$ never leaves $S$).  Let
\begin{equation}
\label{eq1}
E_{t',t}(f,S) = \frac{1}{\mu(S)} \int_{S_{t',t}} G(Df_{t',t}(x)) d\mu(x).
\end{equation}

\textbf{Definition of expansion entropy}:  We define the expansion entropy $H_0$ to be
\begin{equation}
\label{eq2}
H_0(f,S) = \lim_{t'\to\infty} \frac{\ln E_{t',t}(f,S)}{t'-t}.
\end{equation}
We consider $H_0$ and other limiting quantities below to be well-defined only if the limit involved exists.\cite{footnote3}  We remark that if the system $f$ is nonautonomous and the restraining region $S$ is not invariant, $H_0(f,S)$ could potentially depend on the starting time $t$ in addition to $f$ and $S$.  Also, it can be shown that the value of $H_0$ is invariant under differentiable changes of coordinates that are nonsingular on $S$.

To help interpret the definition of $H_0(f,S)$, we now replace $1/\mu(S)$ with $[1/\mu(S_{t',t})][\mu(S_{t',t})/\mu(S)]$ in Eq.~(\ref{eq1}).  The definition (\ref{eq2}) can then be expressed as
\begin{equation}
\label{eq2a}
H_0(f,S) = \lim_{t'\to\infty} \frac{\ln \tilde{E}_{t',t}(f,S)}{t'-t} - \frac{1}{\tau_+},
\end{equation}
where
\begin{equation}
\label{eq2b}
\frac{1}{\tau_+} = \lim_{t'\to\infty} \frac{\ln [\mu(S)/\mu(S_{t',t})]}{t'-t}
\end{equation}
and
$$
\tilde{E}_{t',t}(f,S) = \frac{1}{\mu(S_{t',t})} \int_{S_{t',t}} G(Df_{t',t}(x)) d\mu(x).
$$
Thus, we can view $H_0(f,S)$ as the difference of two exponential rates, given by the limits in Eqs.~(\ref{eq2a}) and (\ref{eq2b}), with the following interpretations.  Imagine that $N$ initial conditions are uniformly sprinkled throughout the volume of $S$ at time $t$, and that $N$ is very large ($N \to \infty$).  The second term $1/\tau_+$ in Eq.~(\ref{eq2a}) is then the exponential decay rate, as $t'$ increases, of the number of trajectories from these initial conditions that remain in $S$ for all times between $t$ and $t'$.  The quantity $\tilde{E}_{t',t}(f,S)$ is the average over these remaining trajectories of the maximum local $d$-dimensional volume growth ratio along the trajectory.  Thus, the first term in Eq.~(\ref{eq2a}) is the exponential growth rate of this average.

It can be shown that the exponential growth rate of $G(Df_{t',t}(x))$ as $t' \to \infty$ is the sum of the positive Lyapunov exponents of the trajectory starting at $x$ at time $t$.  Thus, the limit in Eq.~(\ref{eq2a}) is, in a sense, an average of this sum of positive Lyapunov exponents over trajectories that remain in $S$ for all (forward) time.  The criterion $H_0 > 0$, which we propose for defining chaos, requires that this exponential volume growth rate strictly exceed the exponential rate $1/\tau_+$ at which trajectories leave $S$.  Note that if $S$ is forward invariant, e.g., an absorbing neighborhood of an attractor, then $1/\tau_+ = 0$.

Some points of interest for this entropy definition are that

\begin{enumerate}
\item[(i)] it applies to non-autonomous systems,
\item[(ii)] it assigns an entropy value $H_0$ to every restraining region $S$ in the manifold $M$, and
\item[(iii)] it directly suggests a computational technique for numerically estimating $H_0$ (see Sec.~\ref{sec2.3}).
\end{enumerate}

Finally, notice that
\begin{equation}\label{eqZ}
H_0(f,S') \leq H_0(f,S) \text{ if } S' \subset S.
\end{equation}
This property follows from the fact that $S'_{t',t} \subset S_{t',t}$ if $S' \subset S$, and consequently $E_{t',t}(f,S')  \leq E_{t',t}(f,S)$.  Thus, if there is chaos according to the definition $H_0 > 0$ with respect to a restraining region $S'$, then there is also chaos with respect to every restraining region $S$ that contains $S'$.  In particular, as illustrated by the example in Sec.~\ref{sec2.6.6}, this implies that the expansion entropy will detect chaos ($H_0 > 0$) within a restraining region $S$ when $S$ also contains a nonchaotic attractor, even when the chaos exists only on a repeller.

\subsection{Expansion Entropy of the Inverse System}
\label{sec2.2}

In this section, we show that the expansion entropy of an autonomous, invertible system is the same as the expansion entropy of the inverse system.  (Note that this is also true for the topological entropy; see Sec.~\ref{sec3}.)  This equality results from the following identity for all invertible systems (not necessarily autonomous):
$$
E_{t,t'}(f,S) = E_{t',t}(f,S).
$$
To verify this identity, notice that $f_{t,t'}$ is the inverse of $f_{t',t}$.  Below we use the notation $x' = f_{t',t}(x)$, and consequently $x = f_{t,t'}(x')$.  Then $Df_{t,t'}(x')$ and $Df_{t',t}(x)$ are inverses, and hence the singular values of $Df_{t,t'}(x')$ are the inverses of the singular values of $Df_{t',t}(x)$.  Since the product of the singular values of a square matrix is the absolute value of its determinant, if $A$ is invertible then $|\det A| = G(A)/G(A^{-1})$.  In particular, $|\det Df_{t',t}(x)| = G(Df_{t',t}(x))/G(Df_{t,t'}(x'))$.  Also, $f_{t',t}(S_{t',t}) = S_{t,t'}$.  Writing $E_{t,t'}(f,S)$ as an integral over $x'$ and then making the change of variables $x' = f_{t',t}(x)$,
\begin{eqnarray*}
E_{t,t'}(f,S) &=& \frac{1}{\mu(S)} \int_{S_{t,t'}} G(Df_{t,t'}(x')) d\mu(x') \\
&=& \frac{1}{\mu(S)} \int_{S_{t',t}} G(Df_{t,t'}(x')) |\det Df_{t',t}(x)| d\mu(x) \\
&=& \frac{1}{\mu(S)} \int_{S_{t',t}} G(Df_{t',t}(x)) d\mu(x) = E_{t',t}(f,S).
\end{eqnarray*}

If $f$ is autonomous and invertible, we write $f_{t',t} = f_{t'-t}$ and $f^{-1}_\dt$ = $f_{-\dt}$.  Then
\begin{eqnarray*}
H_0(f^{-1},S)
&=& \lim_{\dt\to\infty} \ln [E_{\dt,0}(f^{-1},S)]/\dt \\
&=& \lim_{\dt\to\infty} \ln [E_{-\dt,0}(f,S)]/\dt \\
&=& \lim_{\dt\to\infty} \ln [E_{0,-\dt}(f,S)]/\dt \\
&=& \lim_{\dt\to\infty} \ln [E_{\dt,0}(f,S)]/\dt = H_0(f,S)
\end{eqnarray*}
Here the first equality is by definition, the second equality is a change of notation, the third equality follows from the time-reversal identity for $E$ derived above, and the fourth equality uses the fact that $f$ is autonomous.

\subsection{Discussion of Numerical Evaluation of Expansion Entropy}
\label{sec2.3}

With respect to point (iii) in Sec.~\ref{sec2.1}, we can imagine a computation of $H_0$ proceeding as follows.  First, randomly sprinkle a large number of initial conditions $\{x_1, x_2, \ldots, x_N\}$ uniformly in $S$.  Then evolve each trajectory $f_{\dt,0}(x_i)$ and the corresponding tangent map $Df_{\dt,0}(x_i)$ forward in time, continuing to evolve only as long as the trajectory remains in $S$.  At a discrete sequence of times $\dt$, compute
\begin{equation}\label{eqW}
\hat{E}_\dt(f,S) = N^{-1} \sum_{i=1}^N \mbox{}' \ G(Df_{\dt,0}(x_i)),
\end{equation}
where the prime on the summation symbol signifies that only those $i$ values for which $f_{\dt,0}(x_i)$ remains in $S$ up to time $t$ are included in the sum.  From our definition of $E$ in Eq.~(\ref{eq1}), we see that $\hat{E}_\dt(f,S)$ is an estimate of $E_{\dt,0}(f,S)$.  Plotting $\ln \hat{E}_\dt(f,S)$ versus $\dt$, for sufficiently large $N$ and $\dt$, we expect to find an approximately linear relationship.  Accordingly, we can estimate $H_0$ as the slope of a straight line fitted to such data (see also Jacobs et al.\cite{JOH} for a similar approach in  two dimensions).

As in other such procedures, judgment and experimentation are called for in determining reliable choices of $N$ and the range of $\dt$ over which to do the fit, and such choices will be constrained by computer resources.  In practice, we find it useful to choose a number, say $100$, of different samples of size $N$, compute $\ln \hat{E}_\dt(f,S)$ for each sample, and take the mean and standard deviations of these logarithms.  Not only does this allow us to estimate the sampling error, it also produces a more reliable mean estimate than computing $\ln\hat{E}_\dt(f,S)$ for a single sample of $100N$ points.  Example illustrations of this computational approach are given in Secs.~\ref{sec2.6.6}, \ref{sec2.6.2}, and \ref{sec2.6.4}.

Specializing to the case of an autonomous invertible system $f$, since Sec.~\ref{sec2.2} shows that the expansion entropy of $f$ and $f^{-1}$ are the same, one could do a numerical computation of $H_0$ using either $f$ or $f^{-1}$.  The question then arises as to which of these two alternatives is preferable from the point of view of computational cost and accuracy.  In the remainder of this section, we argue that it is computationally preferable to calculate $H_0$ from $f$ if $f$ is volume contracting in $S$, while calculation from $f^{-1}$ is preferable if $f$ is volume expanding in $S$.  In order to see this, we generalize Eq.~(\ref{eq2b}) to define both forward and backward exponential decay rates
\begin{equation}
\label{eqX}
\frac{1}{\tau_{\pm}} = \lim_{\dt \to \infty} \frac{1}{\dt} \ln[\mu(S)/\mu(S_{\pm\dt,0})]
\end{equation}
where $S_{\pm\dt,0}$ is, as in Sec.~\ref{sec2.1}, the set of initial conditions at time $0$ whose trajectories under $f$ remain in $S$ between times $0$ and $\pm\dt$, respectively.  That is, in terms of the previously stated numerical procedure for calculating the expansion entropy, $1/\tau_+$ is the exponential temporal decay rate of the number of initial conditions sprinkled uniformly throughout $S$ at time $0$ that lead to orbits that never leave $S$ up to time $\dt$, while $1/\tau_-$ is the analogous quantity taking the initially sprinkled points backward from time $0$ to time $-\dt$.  Since, to estimate the integral in Eq.~(\ref{eq1}), we need to compute the average expansion rates only from those initial conditions that have not left $S$, statistics at any given $\dt$ are improved when the number of such orbits is largest.  Further, the estimate of the limit $\dt \to +\infty$ dictates that we make $\dt$ large.  These two considerations indicate that the forward (respectively, backward) calculation of $H_0$ will be computationally more efficient if $\tau_+ > \tau_-$ (respectively, $\tau_- > \tau_+$).  Subtracting the definition (\ref{eqX}) of $1/\tau_+$ from the definition of $1/\tau_-$, and using the fact that $S_{-\dt,0} = f_{\dt,0}(S_{\dt,0})$ for an autonomous invertible system, we obtain
$$
(1/\tau_+) - (1/\tau_-) = \lim_{\dt \to \infty} \frac{1}{\dt} \ln\{\mu(f_{\dt,0}(S_{\dt,0}))/\mu(S_{\dt,0})\}.
$$
The right hand side is positive (respectively, negative) when the map is volume expanding (respectively, contracting) in $S$.
Thus, $\tau_- > \tau_+$ if the map is volume expanding, while $\tau_+ > \tau_-$ if the map is volume contracting.  In particular, if $S$ is a neighborhood of an attractor, it is best to employ a forward time calculation.  We note that the common examples of the H\'enon map and the Lorenz system are uniformly volume contracting at all points in state space (implying that $\tau^+ > \tau^-$), while Hamiltonian systems are volume preserving (implying that $\tau^+ = \tau^-$).

\subsection{Generalization to $q$-order expansion entropy}
\label{sec2.5}

In past work on fractal dimension, the box-counting dimension has been generalized to a spectrum of dimensions often denoted $D_q$, where the box-counting dimension corresponds to $q=0$, and the index $q$ can be any nonnegative number\cite{BR,HP,GP2}.  In addition, a spectrum of entropy-like quantities, again depending on an index $q \geq 0$, has been introduced by Grassberger and Procaccia\cite{GP1,GP2}, where $q=0$ corresponds to the topological entropy, and $q=1$ corresponds to the metric entropy.  Thus, motivated by these past works, it is natural to introduce an analogous spectrum of $q$-order expansion entropies, $H_q$, and to consider whether they are useful with respect to the issue of defining chaos.

In Appendix~\ref{appA}, we introduce and discuss a natural way of specifying $H_q$.  In particular, the form defining $H_q$ is specified so that it gives Eqs.~(\ref{eq1}) and (\ref{eq2}) when $q=0$, gives an expansion entropy analogue of the entropy of Grassberger and Procaccia\cite{GP1,GP2}, and also gives a correspondence for $q=1$ with previous results for the metric entropy of repellers\cite{KG,HOY} and with Pesin's formula\cite{P} for the metric entropy for attractors.  However, as we will argue in Appendix~\ref{appA}, $q=0$ is special in regard to defining chaos.  In particular, Appendix~\ref{appA} will consider $H_q$ for an example in which $S$ contains an attracting fixed point and a chaotic repeller (see Sec.~\ref{sec2.6.6}).  For this example, it is shown that $H_0 > 0$, while $H_q$ for $q > 0$ can be zero.  Thus, $H_0$ successfully detects the chaos within $S$, but $H_q$ for $q > 0$ may not.

\section{Illustrative Examples}
\label{sec2.6}

\subsection{Attracting and repelling fixed points}
\label{sec2.6.1}

Consider a one-dimensional differentiable map $f$ with a fixed point $x_0$; for simplicity, we assume that $Df(x_0) \neq \pm 1$.  Let the restraining region $S$ be an interval containing $x_0$ on which $|Df(x)| \neq 1$; that is, $f$ is either uniformly expanding or uniformly contracting on $S$.  In either case, we show below that the expansion entropy $H_0(f,S)$ is zero, i.e., that fixed points are not chaotic according to our definition.

In the case of an attracting fixed point, $|Df_{t',t}(x)| < 1$, and hence $G(Df_{t',t}(x)) = 1$, for all $x \in S$ and $t' > t$.  Also, $S_{t',t} = S$ for $t' > t$.
Then from Eq.~(\ref{eq1}) and (\ref{eq2}) we have $E_{t',t}(f,S) = \mu(S)$ for $t' > t$ and $H_0(f,S) = 0$.

In the expanding case, $|Df_{t',t}(x)| > 1$ for trajectories that remain in $S$ from time $t$ to $t'$.  Also, $S_{t',t}$ is a subinterval of $S$ whose endpoints map to the endpoints of $S$ under $f_{t',t}$.  Thus,
\begin{eqnarray*}
E_{t',t}(f,S)
&=& \frac{1}{\mu(S)} \int_{S_{t',t}} |Df_{t',t}(x)| dx \\
&=& \frac{1}{\mu(S)} \left|\int_{S_{t',t}} Df_{t',t}(x) dx\right| = 1.
\end{eqnarray*}
Once again, $H_0(f,S) = 0$.

For fixed points (or periodic orbits) of higher-dimensional systems, similar arguments can be made, though they are more complicated in the case when the fixed point has both stable and unstable directions.  The essence of these calculations is that any growth in the integrand $G$ of Eq.~(\ref{eq1}) as $t' - t$ increases is balanced (up to a time-independent multiplicative constant) by a reduction in the volume of $S_{t',t}$.  The conclusion remains that $H_0 = 0$, i.e., that isolated period orbits are not chaotic.

\subsection{Example: A one-dimensional map with a chaotic repeller and an attracting fixed point}
\label{sec2.6.6}

\begin{figure}
\includegraphics[width=3.25in]{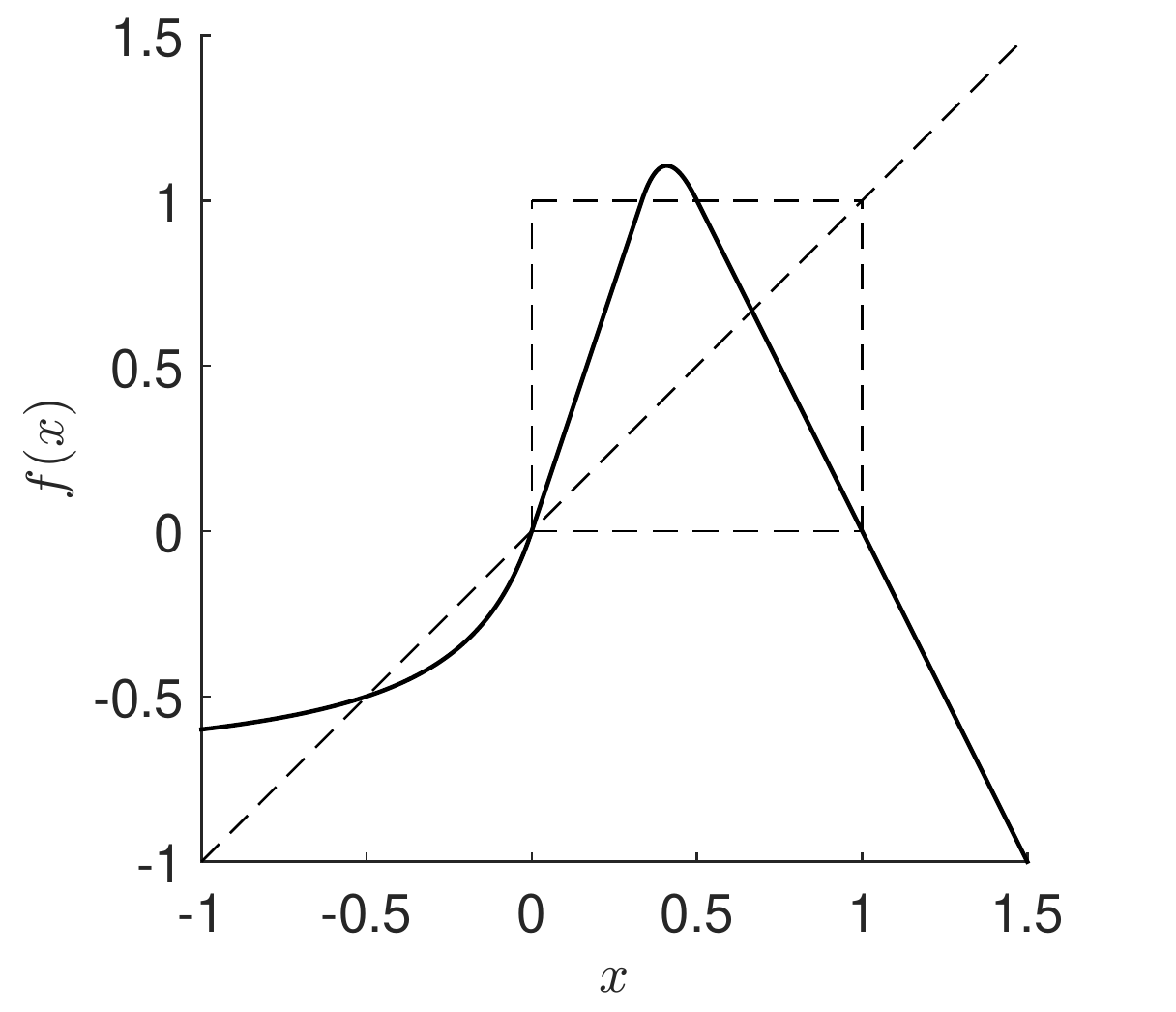}
\caption{\label{fig1d}A map illustrating a case where $S = [-1,1.5]$ contains a chaotic repeller (in $S' = [0,1]$) and an attracting fixed point ($x = -1/2$).}
\end{figure}

Consider the one-dimensional map $f$ shown in Fig.~\ref{fig1d} and the two restraining regions $S$ and $S' \in S$, where we take $S = [-1,1.5]$ and $S' = [0,1]$.  The map is linear with derivative $3$ on $[0,1/3]$ and linear with derivative $-2$ on $[1/2,1]$, mapping each of these intervals to $[0,1]$.
For this example, the fixed point $x = -1/2$ attracts almost all initial conditions with respect to Lebesgue measure in $S$.  

On the other hand, $S'$ contains an invariant Cantor set that is commonly called a chaotic repeller.  We now show that according to our definition, $f$ exhibits chaos by having positive expansion entropy $H_0$ in both $S$ and $S'$.  The invariant Cantor set consists of all initial conditions in $[0,1]$ whose trajectories never land in the interval $(1/3,1/2)$, i.e., those whose base $6$ expansion does not contain the digit $2$.  The set $S'_{\dt,0}$ of initial conditions whose trajectories remain in $[0,1]$ from time $0$ to time $\dt$ consists of $2^\dt$ intervals corresponding to all possible strings of length $\dt$ of the letters $L$ and $R$, where $L$ denotes an iteration in the interval $[0,1/3]$ and $R$ denotes an iteration in the interval $[1/2,1]$.  A string with $k$ $L$'s and $\dt-k$ $R$'s corresponds to an interval of length $3^{-k} 2^{k-\dt}$, on which $Df^\dt = 3^k (-2)^{\dt-k}$.  Then the integral of $G(Df^\dt) = |Df^\dt|$ on each such interval is $1$.  Thus, $E_{\dt,0}(f,S') = 2^\dt$, and hence $H_0(f,S') = \ln 2$.
In accordance with property (\ref{eqZ}), since $S$ contains $S'$, we also have $H_0(f,S) = \ln 2$.

In Appendix~\ref{appA}, in addition to defining the quantity $H_q$ discussed in Sec.~\ref{sec2.5}, we also evaluate $H_q$ for the map in Fig.~\ref{fig1d}.  We find for the smaller restraining region $S'$ that $H_q(f,S') > 0$ for all $q \geq 0$, but for the larger restraining region $S$ there is a critical value $q_c < 1$ such that $H_q(f,S) = 0$ for $q \geq q_c$.  For $q=1$, we have $H_1(f,S') = (2/5) \ln (5/2) + (3/5) \ln (5/3) > 0$, while $H_1(f,S) = 0$.  Our interpretation is that $H_1(f,S)$ is dominated by the dynamics of Lebesgue almost every initial condition whose trajectory approaches the fixed point attractor, while $H_0(f,S)$ is dominated by the chaotic saddle in $S'$.  We therefore conclude that $H_1$ (and similarly $H_q$ for $q > 0$) is not an appropriate tool for detecting non-attracting chaos in a restraining region.

\begin{figure}
\includegraphics[width=3.25in]{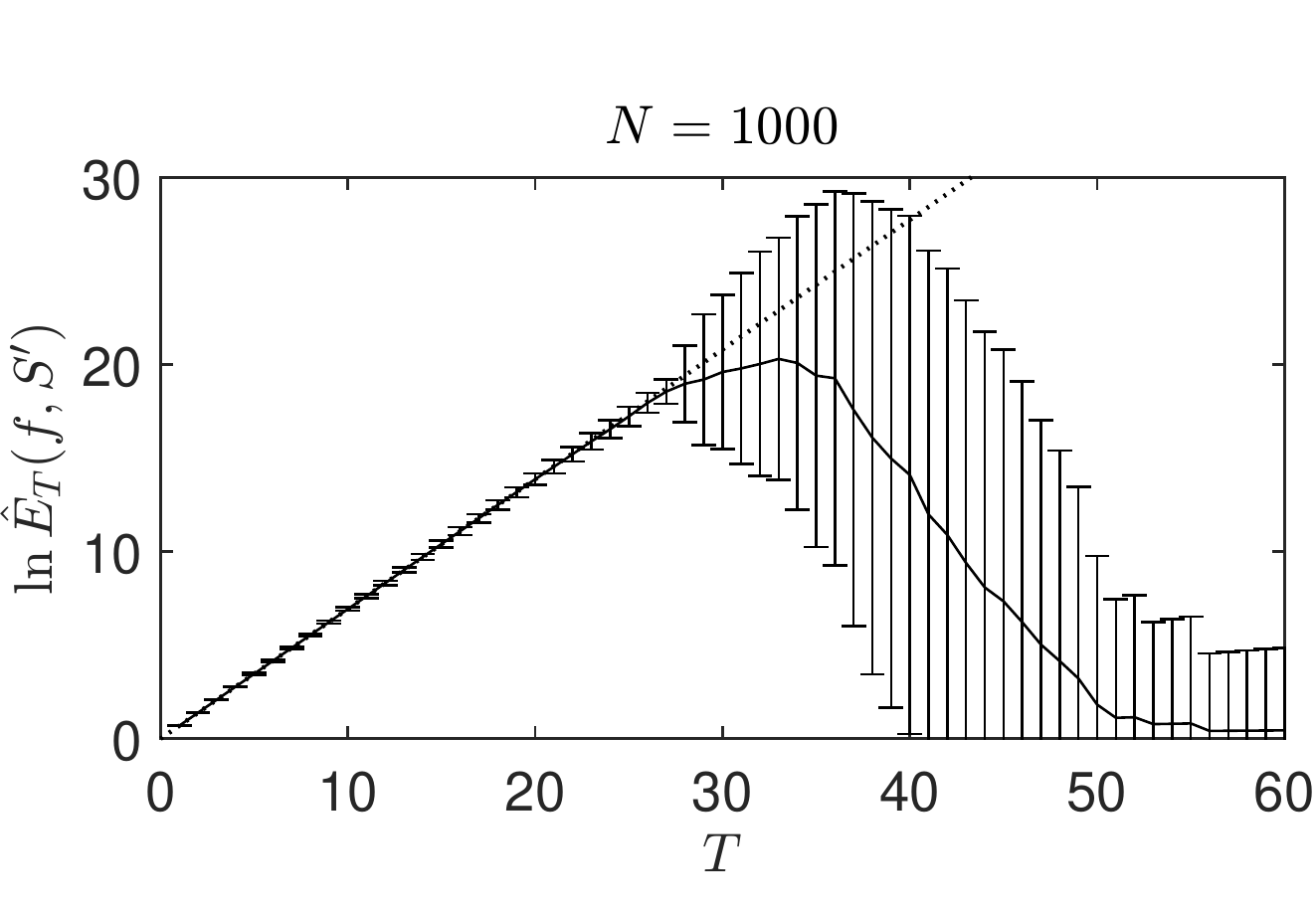}
\includegraphics[width=3.25in]{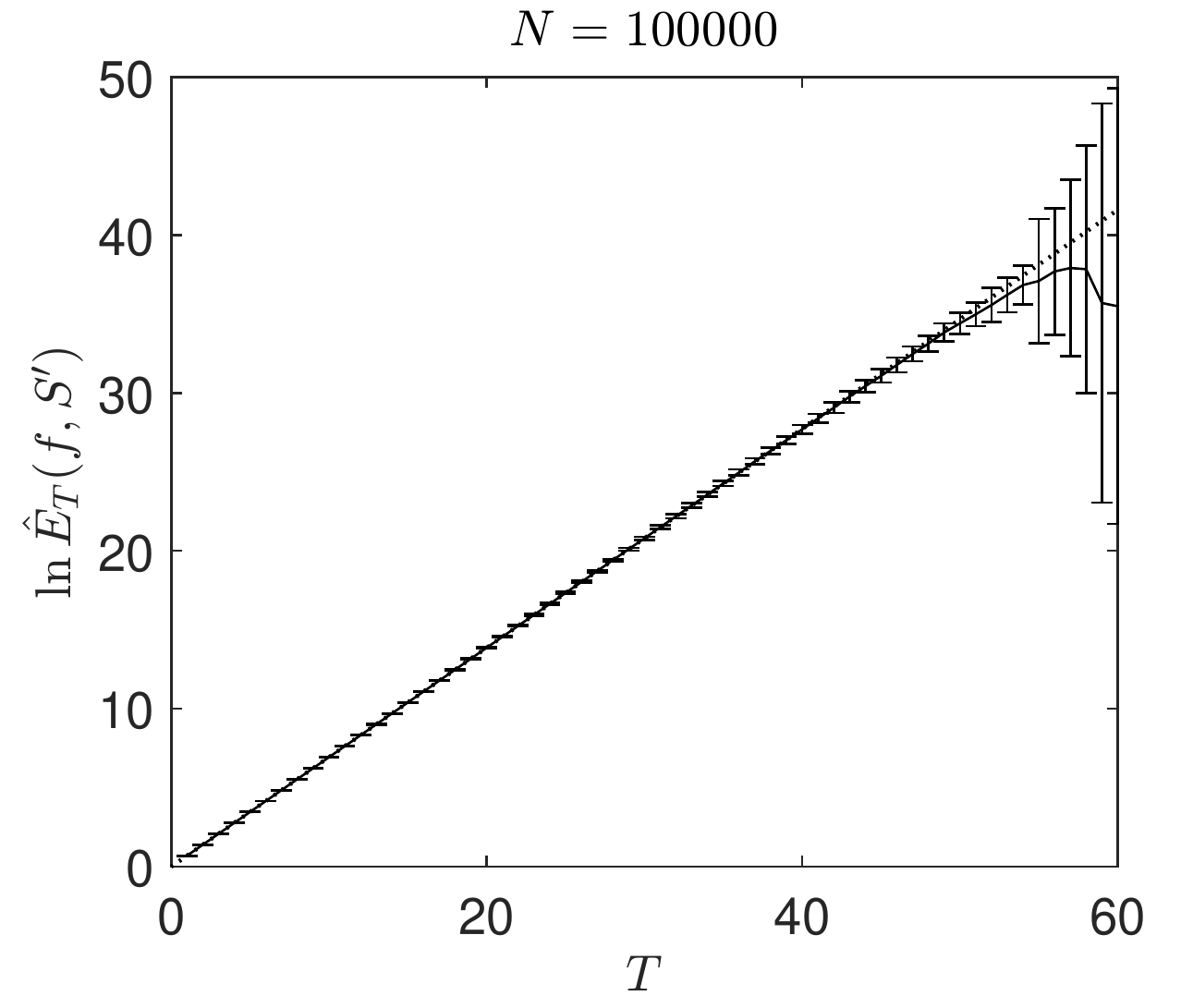}
\caption{\label{fig1d1}Computation of $\ln \hat{E}_T$ versus $T$ for the map of Fig.~\ref{fig1d} with restraining region $S'$.  For each $T$, we computed $\ln\hat{E}_T(f,S')$ for $100$ different samples, with $N=1000$ randomly chosen initial conditions in each sample (top figure) and $N=100,000$ randomly chosen initial conditions in each samples (bottom figure).  The solid curve shows the mean of the $100$ samples, and the error bars show their standard deviation.
As we discussed in Sec.~\ref{sec2.3}, the slope of the solid curve should, in the limit of large $N$ and $T$, approximate $H_0(f,S')$.  The dashed line has slope $\ln 2$, which is the value we obtained analytically for $H_0(f,S')$.}
\end{figure}

\begin{figure}
\includegraphics[width=3.25in]{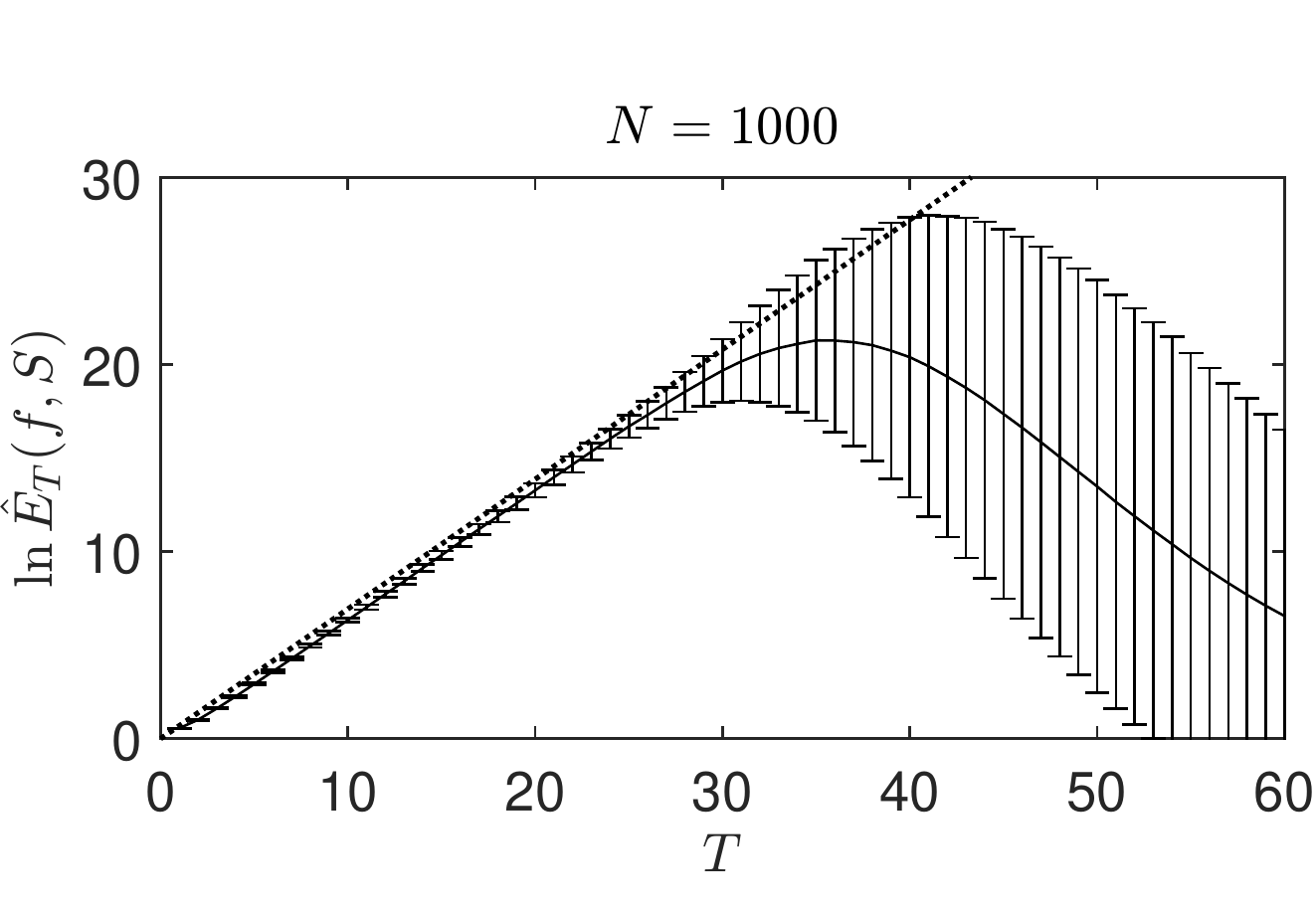}
\includegraphics[width=3.25in]{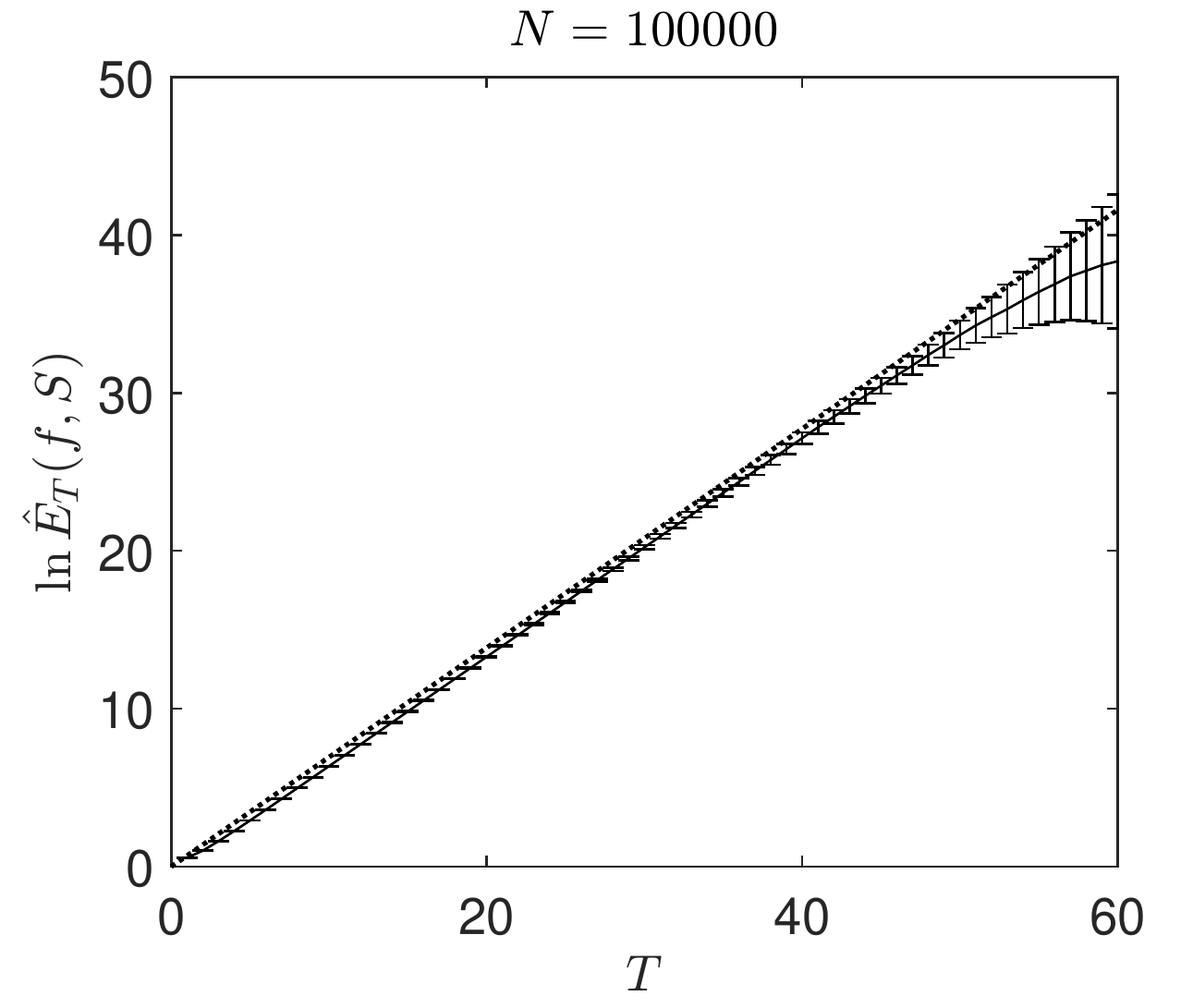}
\caption{\label{fig1d2}Computation of $\ln \hat{E}_T$ versus $T$ for the map of Fig.~\ref{fig1d} with restraining region $S$.  This is the analogue of Fig.~\ref{fig1d1} with $S'$ replaced by $S$.}
\end{figure}

Next we use this example to illustrate the numerical computation of $H_0$.  Our procedure, as explained in Sec.~\ref{sec2.3}, is to choose a sample size $N$ and range of $T$ values and to do the following for each $T$.  Using Eq.~(\ref{eqW}), we compute an estimate $\hat{E}_T$ of $E_{T,0}$ for each of $100$ different samples of $N$ points each in the restraining region, and compute the mean and standard deviation of the $100$ samples.  The results for $N = 1000$ and $N = 100,000$ are shown in Fig.~\ref{fig1d1} for $S'$ and in Fig.~\ref{fig1d2} for $S$.  The estimated value of $H_0$ is the slope of the solid curve in an appropriate scaling interval.  The scaling interval for a given $N$ can be judged by consistency of the results with a larger value of $N$, in addition to smallness of the error bars and straightness of the curve.  Notice that the somewhat arbitrarily chosen restraining region $S$ yields nearly as long a scaling interval as the restraining region $S'$ that is chosen with knowledge of the invariant Cantor set.

\subsection{Example: A random one-dimensional map}
\label{sec2.6.2}

Consider the one-dimensional random map
\begin{equation}
\label{eqK}
\theta_{t+1} = [\theta_t + \alpha_t + K \sin\theta_t] \text{ mod } 2\pi
\end{equation}
where $K > 0$ and $\alpha_0, \alpha_1, \alpha_2, \ldots$ are independent random variables that are uniformly distributed in the circle $[0,2\pi)$.  We take the restraining region $S$ to be the entire circle.

Notice that
$$\left| \frac{d\theta_\dt}{d\theta_0} \right| = \prod_{t = 0}^{\dt-1} |1 + K\cos\theta_t|,
$$
and that
$$
E_{\dt,0} = \langle \max(|d\theta_T/d\theta_0|,1) \rangle_{\theta_0},
$$
where $\langle\cdots\rangle_\eta$ denotes an average over $\eta$.  If $\theta_0$ is uniformly distributed, then $\theta_0, \theta_1, \theta_2, \ldots$ are \emph{independent} and uniformly distributed, so that
\begin{eqnarray*}
\langle |d\theta_\dt/d\theta_0| \rangle_{\theta_0,\theta_1,\ldots,\theta_{\dt-1}}
&=& \prod_{t = 0}^{\dt-1} \langle |1 + K\cos\theta_t| \rangle_{\theta_t} \\
&=& \langle |1 + K\cos\theta| \rangle_{\theta}^\dt
\end{eqnarray*}
This suggests that for a typical realization of the random inputs, $H_0 \approx \lambda$, where
$$
\lambda = \ln \langle |1 + K\cos\theta| \rangle_{\theta}.
$$
For $0 < K \leq 1$,
$$
\lambda = \ln \langle (1 + K\cos\theta) \rangle_{\theta} = \ln 1 = 0,
$$
while for $K > 1$,
$$
\lambda = \ln \langle |1 + K\cos\theta| - (1 + K\cos\theta) + 1 \rangle_\theta > \ln 1 = 0.
$$

\begin{figure}
\includegraphics[width=3.25in]{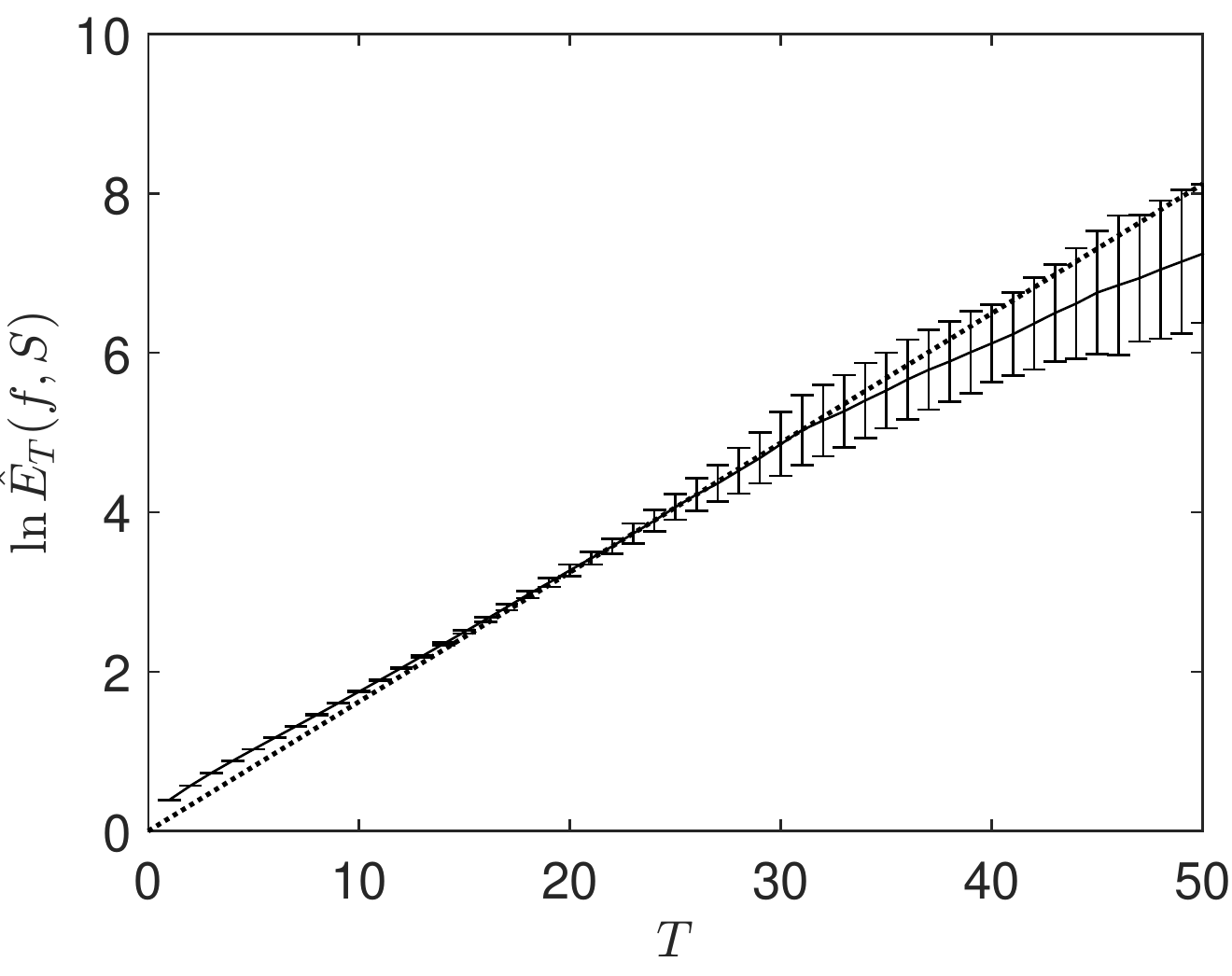}
\caption{\label{fig0}Computation of $\ln \hat{E}_T$ versus $T$ for the random circle map (\ref{eqK}) with $K = 1.5$.
For each $T$, we computed $\ln\hat{E}_T(f,S)$ for $100$ different samples, with $N = 1,000,000$ randomly chosen initial conditions in each sample.
Each initial condition used a different realization of the random sequence of maps.
The solid curve shows the mean of the $100$ samples, and the error bars show their standard deviation.  The dashed line has slope $\ln \langle |1 + 1.5 \cos\theta| \rangle_\theta$; this estimate of the expansion entropy appears to be slightly larger than the slope of the computed data.}
\end{figure}

For $0 < K \leq 1$, each map is a diffeomorphism, so $d\theta_t/d\theta_0 > 0$, and
$$
E_{\dt,0} = \langle\max(d\theta_\dt/d\theta_0,1)\rangle_\theta < \langle d\theta_\dt/d\theta_0 + 1 \rangle_{\theta_0} = 2.
$$
Thus, $E_{\dt,0}$ is not exponentially increasing, so indeed $H_0 = 0$ for $0 < K \leq 1$.  Numerical experiments agree with the argument above that $H_0 > 0$ for $K > 1$, though establishing that the transition to chaos (according to our definition) occurs exactly at $K = 1$ would require a more definitive study.  In Fig.~\ref{fig0}, we show the computed $\ln \hat{E}_T$ (see Sec.~\ref{sec2.3} and \ref{sec2.6.6}) versus $T$ for $K = 1.5$.

\subsection{Example: Shear map on the 2-torus}
\label{sec2.6.3}

\begin{figure}
\includegraphics[width=3.25in]{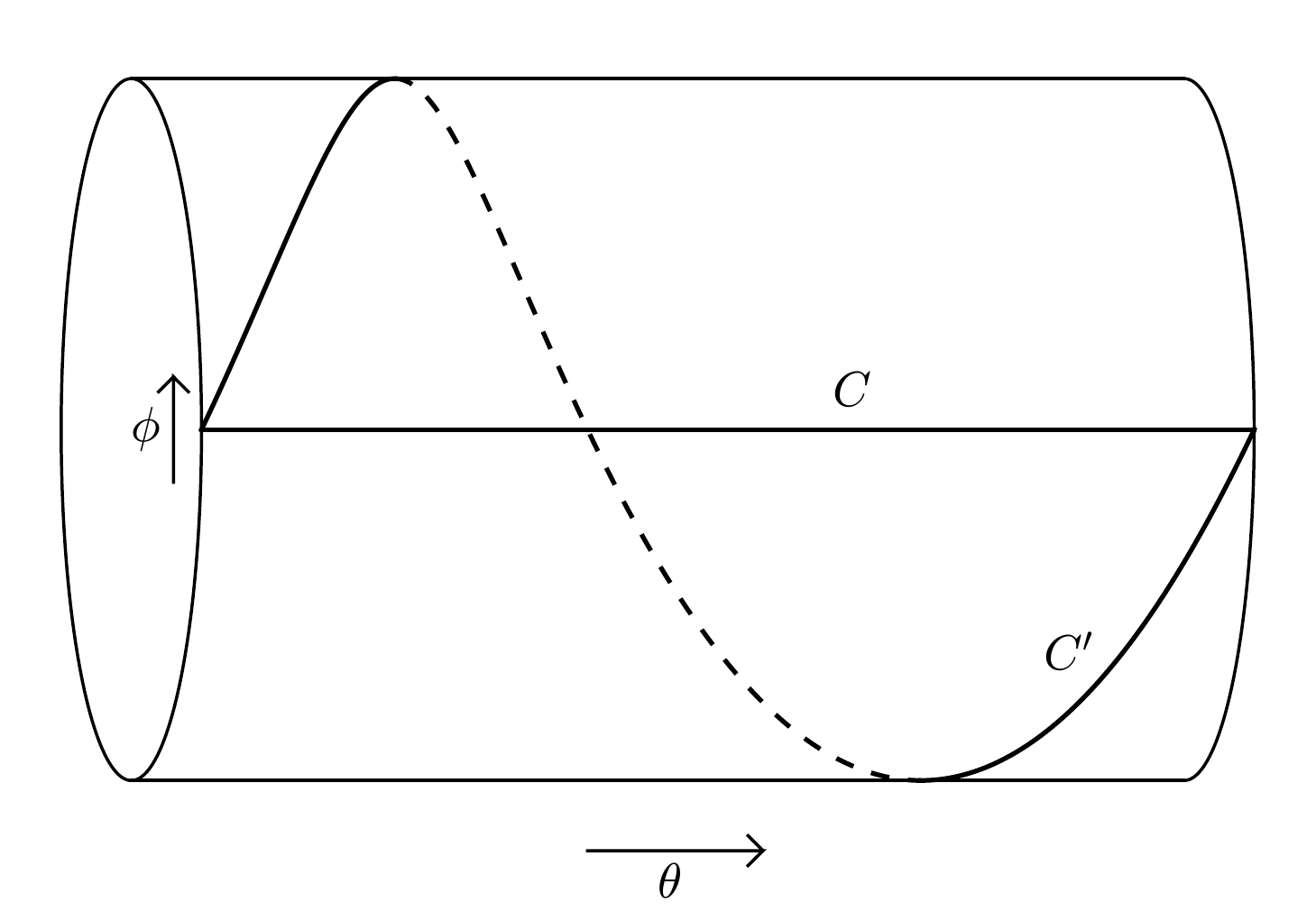}
\caption{\label{fig1}Image $C'$ of a circle $C$ given by $\phi = \text{constant}$ under the shear map (\ref{eqY}).}
\end{figure}

This example illustrates a case where orbits are dense and, as in chaos, typical nearby orbits separate from each other with increasing time.  However, the rate of separation is linear, rather than exponential in time.  The example is the following map of the 2-torus:
\begin{equation}
\label{eqY}
\phi_{t+1} = [\phi_t + \theta_t] \text{ mod } 2\pi, \quad \theta_{t+1} = [\theta_t + \omega] \text{ mod } 2\pi,
\end{equation}
where $\omega/(2\pi)$ is irrational\cite{R,HO}.  As shown in Fig.~\ref{fig1}, the image under this map of a curve $C$ looping around the $\theta,\phi$-torus once in the $\theta$ direction is  a curve $C'$ that loops once in the $\phi$ direction, as well as once in the $\theta$ direction, with the number of $\phi$ loops increasing by one on each subsequent iterate.  Orbits are dense and nearby initial conditions with different values of $\theta$ separate linearly with $t$.  To evaluate $H_0$ for this map from the definition Eq.~(\ref{eq2}), with the restraining region $S$ being the entire torus, we note that
$$
Df_{t+1,t} = \left(\begin{array}{cc}1 & 1\\ 0 & 1\end{array}\right) \text{ and } Df_{t',t} = \left(\begin{array}{cc}1 & t' - t\\ 0 & 1\end{array}\right)
$$
for all $\phi$ and $\theta$.  For $t' - t \gg 1$, the singular values of $Df_{t',t}$ are approximately $t' - t$ and $(t' - t)^{-1}$.  Thus, for large $t' - t$,
$$
E_{t',t}(f,S) \approx (2\pi)^2 (t' - t),
$$
and $H_0 = 0$ (since $(t' - t)^{-1} \log(t' - t) \to 0$ as $t' \to \infty$).  Hence this example is not chaotic according to our definition.

\subsection{Example: Horseshoe and H\'enon map}
\label{sec2.6.4}

\begin{figure}
\includegraphics[width=3.25in]{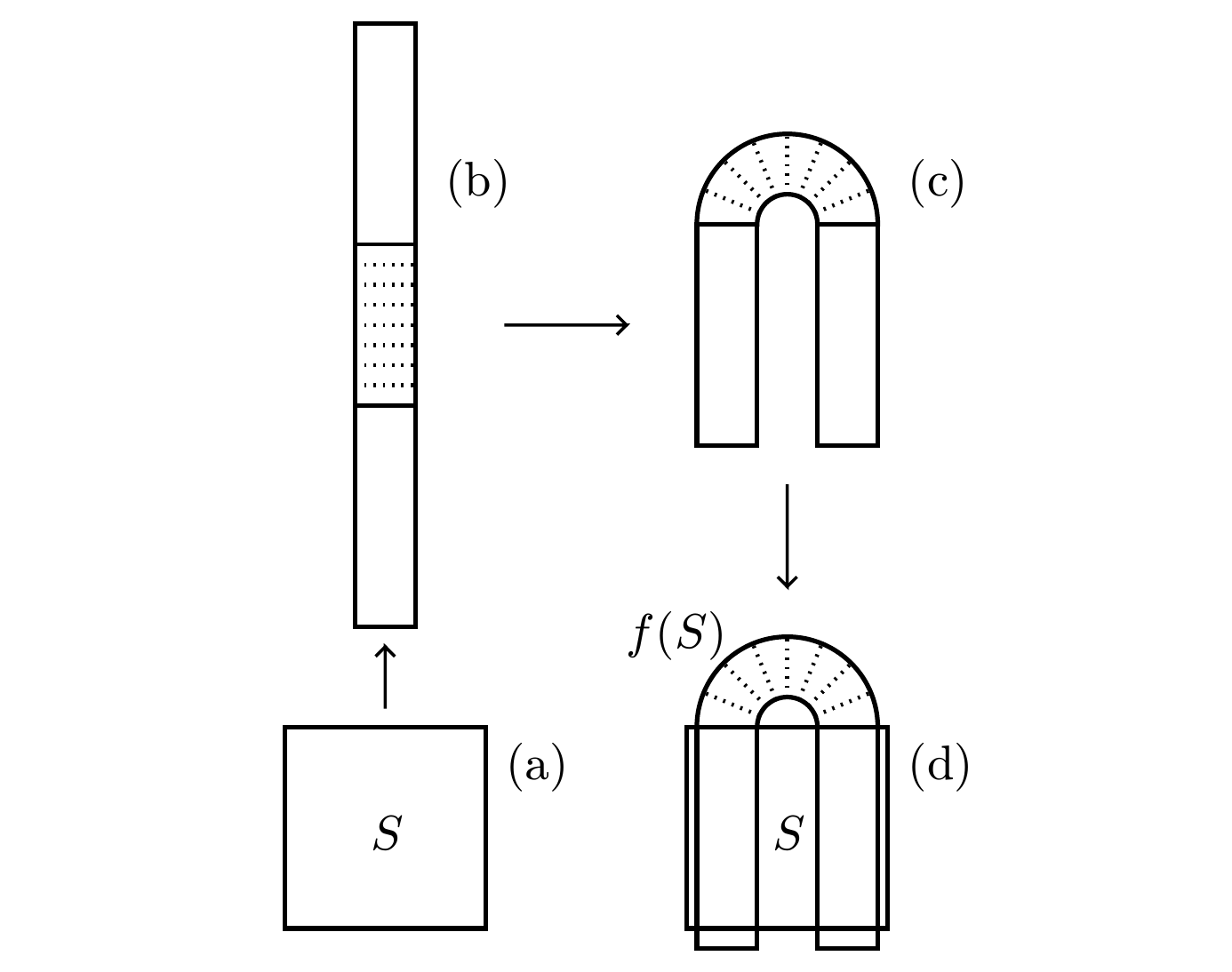}
\caption{\label{fig2}A horseshoe map $f$ that is linear for trajectories that remain in the restraining region $S$.}
\end{figure}

Figure~\ref{fig2} shows the action of a horseshoe map in the plane on a unit square $S$, which we also take to be the restraining region.  The step (a) $\to$ (b) represents a uniform horizontal compression and vertical stretching of the square.  Let $\rho > 2$ be the ratio by which the vertical length of the square is stretched, and assume that bending deformations in the step (b) $\to$ (c) take place only in the shaded region.  The fraction of the original square that remains in the square after one iterate is $2/\rho$ (see Fig.~\ref{fig2}(d)), and after $t' - t$ iterates, the fraction is $(2/\rho)^{t' - t}$.  Also, $G(Df_{t',t}) = \rho^{t' - t}$, yielding $E_{t',t}(f,S) = \rho^{t' - t} (2/\rho)^{t' - t} = 2^{t' - t}$, and $H_0 = \ln 2$.  Thus, by our definition the horseshoe map is chaotic in $S$.

\begin{figure}
\includegraphics[width=3.25in]{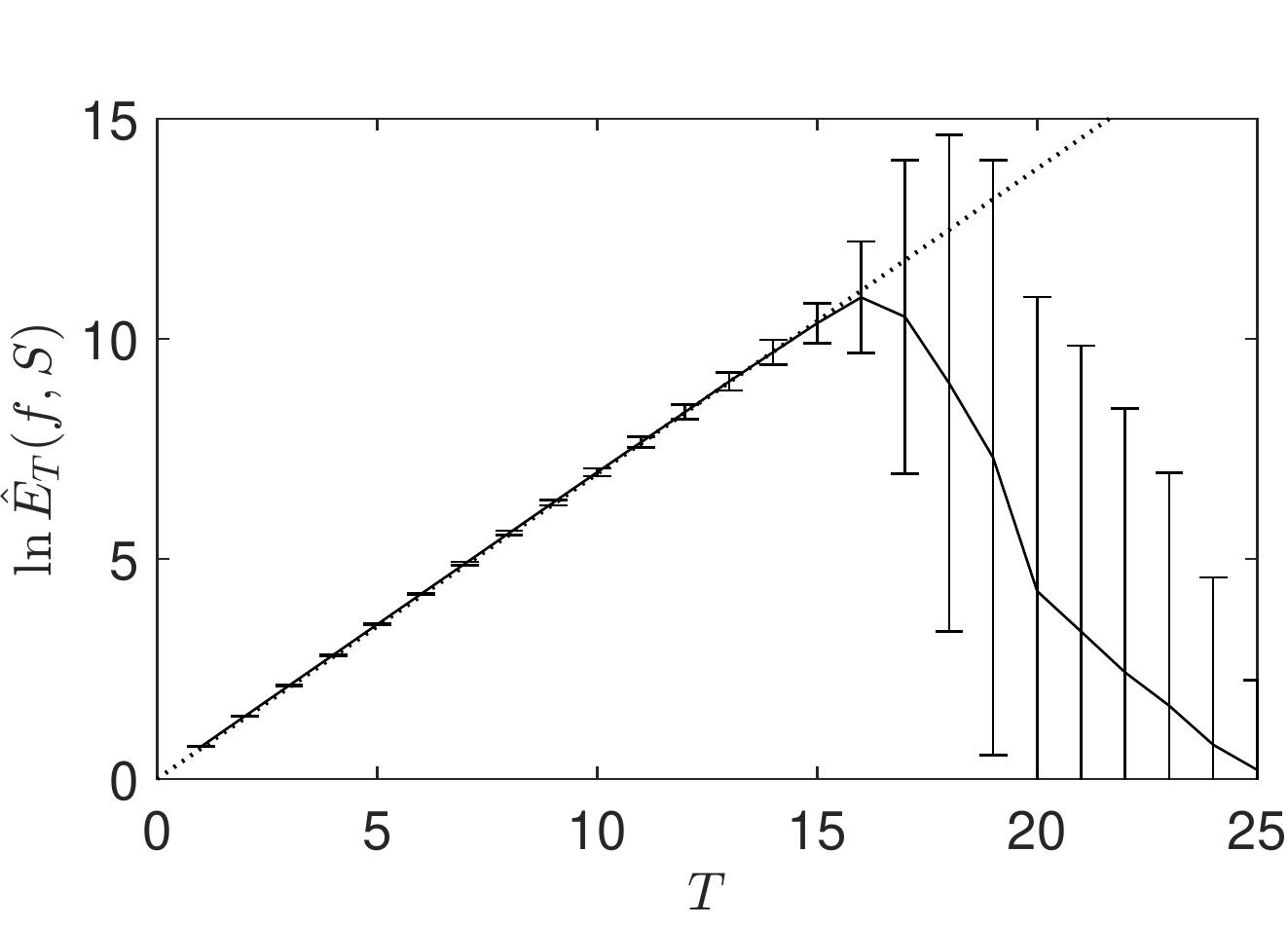}
\caption{\label{fig2n}Computation of $\ln \hat{E}_T$ versus $T$ for the H\'enon map with $a = 4.2$ and $b = 0.3$.  For each $T$, we computed $\ln\hat{E}_T(f,S)$ for $100$ different samples, with $N = 100,000$ randomly chosen initial conditions in each sample.  The solid curve shows the mean of the $100$ samples, and the error bars show their standard deviation.  The dashed line has slope $H_0 = \ln 2$.}
\end{figure}

For the H\'enon map
$$
x_{t+1} = a + b y_t - x_t^2, \quad y_{t+1} = x_t
$$
with $b = 0.3$, the results of Devaney and Nitecki\cite{DN} imply that for $a \geq 3.4$, the map has a topological horseshoe, and
for $a \leq 5.1$, the nonwandering set is contained in the square $-3 \leq x,y \leq 3$, which we take to be the restraining region $S$.  Fig.~\ref{fig2n} shows the results of a numerical computation for $a = 4.2$ of $\ln \hat{E}_T$ (see Sec.~\ref{sec2.3} and \ref{sec2.6.6}) versus $T$, which agrees well with the value $H_0 = \ln 2$.

\section{Topological Entropy}
\label{sec3}

In this section, we define \emph{topological entropy}, and discuss its relation to (and equivalence with, in appropriate circumstances) both expansion entropy and the related notion of \emph{volume growth}.

The original definition of topological entropy, by Adler, Konheim, and McAndrew\cite{AKM}, was for a continuous map $f$ on a compact topological space $X$.   
If $X$ is a metric space, an equivalent definition of topological entropy due to Dinaburg\cite{Din} and Bowen\cite{Bo1} is as follows.  (Equivalence to the original definition was proved by Bowen\cite{Bo2}.)  For $\epsilon > 0$, two points $x$ and $y$ in $X$ are called $(\dt,\epsilon)$-separated if the distance between their $k$th iterates satisfies $d(f^k(x),f^k(y)) > \epsilon$ for some $0 \leq k < \dt$.  A finite set of points $P \subset X$ is said to $(\dt,\epsilon)$-span $X$ if there is no point in $X$ that is $(\dt,\epsilon)$-separated from every point in $P$.  Let $n(\dt,\epsilon)$ be the minimum number of points needed to $(\dt,\epsilon)$-span $X$, and let $N(\dt,\epsilon)$ be the maximum number of points in $X$ that can be pairwise $(\dt,\epsilon)$-separated.  Let
$$
h_n(f,\epsilon) = \limsup_{\dt\to\infty} \frac{\ln n(\dt,\epsilon)}{\dt}
$$
and
\begin{equation}\label{eq00}
h_N(f,\epsilon) = \limsup_{\dt\to\infty} \frac{\ln N(\dt,\epsilon)}{\dt}.
\end{equation}
It is not hard to show that $n(\dt,\epsilon) \leq N(\dt,\epsilon) \leq n(\dt,\epsilon/2)$.  This implies the analogous relation between $h_n$ and $h_N$, which implies that they have the same limit as $\epsilon \to 0$.  Define the topological entropy $h$ of $f$ on $X$ by
\begin{equation}\label{eq0}
h(f,X) = \lim_{\epsilon\to 0} h_n(f,\epsilon) = \lim_{\epsilon\to 0} h_N(f,\epsilon).
\end{equation}
 
The notions of expansion entropy $H_0$ and topological entropy $h$ are both well-defined in the case when $f$ is a smooth, autonomous system on a compact manifold $M$ and the restraining region $S$ is all of $M$.  In this case, Sacksteder and Shub\cite{SS} defined a quantity they called $h_1$ that is equivalent to expansion entropy.  Subsequently, Przytycki\cite{Pr} proved that $h_1$ is an upper bound on $h$ if $f$ is a $C^{1+\gamma}$ diffeomorphism for $\gamma > 0$; this proof was extended to noninvertible maps by Newhouse\cite{N}.
Though there are examples\cite{MS} for which the two quantities differ, Kozlovski\cite{K} proved that $h_1 = h$ for $C^\infty$ maps.  Thus, $H_0(f,M) = h(f,M)$ for a sufficiently smooth map $f$ on a compact manifold $M$.

From our point of view, these results leave open consideration of important issues regarding nonautonomous systems and the role of restraining regions.  For example, suppose now that $J$ is a compact invariant set of an autonomous system $f$ on a (not necessarily compact) manifold $M$.  If $J$ has volume zero, $H_0(f,J)$ is undefined, but we can define $H_0$ for a neighborhood $S$ of $J$ that contains no other invariant sets.  In this case, we conjecture that $H_0(f,S) = h(f,J)$ if $f$ is $C^\infty$.  More generally, when the restraining region $S$ contains multiple invariant sets, we conjecture (consistent with Eq.~(\ref{eqZ})) that $H_0(f,S)$ is the maximum topological entropy of $f$ on an invariant subset of $S$.

Our notion of expansion entropy is related  to the notion of volume growth defined by Yomdin\cite{Y} and Newhouse\cite{N}.  Yomdin defines the exponential rate $v_d(f)$ of $d$-dimensional volume growth of a smooth map $f$ on a compact manifold $M$, and proves that $v_d(f) \leq h(f,M)$ if $f$ is $C^\infty$.  Newhouse defines the volume growth rate more generally for a neighborhood $U$ of a compact invariant set $J \subset M$, and proves that $h(f,J)$ is bounded above by the maximum over $d$ of the $d$-dimensional volume growth rate on $U$.  See also Gromov\cite{G} for a discussion of these results.

Based on these results, Newhouse and Pignataro\cite{NP} proposed and implemented algorithms for computing entropy of two-dimensional diffeomorphisms (including Poincar\'e sections of three-dimensional differential equations) by computing the exponential growth rate of the length of an iterated curve.
Other algorithms\cite{KT,COH} compute the growth rate of the number of disconnected arcs
resulting from the iteration of an initial line segment
within a neighborhood of a two-dimensional chaotic saddle or repeller.  Of course, these methods could be extended to higher dimensions by considering growth of surface areas, etc.  Expansion entropy, by estimating volume growth locally, allows an analogous computation to be done without having to compute and measure multidimensional surfaces.
It is analogous to the approach used by Jacobs et al.\cite{JOH} for two-dimensional maps.

Another approach to computing entropy is by symbolic dynamics: partition the state space into numbered subsets, and estimate the exponential growth rate (as time increases) of the number of different sequences of subsets that can be visited by a finite-time trajectory.  This approach can yield good estimates with well-chosen partitions, but inadequate partitions may lead to underestimation, and in some cases symbolic dynamics indicates positive entropy when the topological entropy is actually zero.\cite{FJO} 

We conclude this section with a brief discussion of the connection between the definitions of expansion entropy and topological entropy in the case of a smooth, autonomous system on a compact manifold $M$, with restraining region $S=M$.  In Appendix~\ref{appB}, we argue that for sufficiently small $\epsilon > 0$, the quantity $E_{T,0}(f,S)$ of Eq.~(\ref{eq1}) approximates $\tilde{N}(T,\epsilon)/N(0,\epsilon)$, where $\tilde{N}(T,\epsilon)$ is the maximum number of trajectories that are a distance $\epsilon$ apart at either time $0$ or at time $T$.  Note that $\tilde{N}(T,\epsilon)$ is a lower bound on $N(T,\epsilon)$, because the latter distinguishes between trajectories that are $\epsilon$ apart at some time between $0$ and $T$; however, at least for hyperbolic systems the difference between $\tilde{N}$ and $N$ should be inconsequential.
Eqs.~(\ref{eq00}) and (\ref{eq0}) first take a limit with respect to $T$ and then $\epsilon$.  Normalizing by $N(0,\epsilon)$ does not change the limit:
\begin{eqnarray}\label{eqA}
h(f,S)
&=& \lim_{\epsilon \to 0} \limsup_{T\to\infty} \frac{\ln N(T,\epsilon) - \ln N(0,\epsilon)}{T} \nonumber\\
&=& \lim_{\epsilon \to 0} \limsup_{T\to\infty} \frac{\ln (N(T,\epsilon)/N(0,\epsilon))}{T}.
\end{eqnarray}
We have argued above that
$$
E_{T,0}(f,S) = \lim_{\epsilon \to 0} \frac{\tilde{N}(T,\epsilon)}{N(0,\epsilon)}.
$$
Thus, by Eq.~(\ref{eq2}), the definition of $H_0$ differs from the definition of $h$ primarily because it uses $\tilde{N}(T,\epsilon) \leq N(T,\epsilon)$, and because the limits with respect to $T$ an $\epsilon$ are taken in the reverse order.

\section{Definitions of Chaos that do not Involve Entropy: Sensitive Dependence, Lyapunov Exponents and Chaotic Attractors}
\label{sec4}

A concept often associated with chaos is that of sensitive dependence, the idea that the orbits from two nearby initial conditions can move far apart in the future.  In the mathematical literature the most common definition of ``sensitive dependence'' is as follows.  (This definition is also sometimes called  ``weak sensitive dependence.'')

\textbf{Definition}: A continuous map, $f:M\to M$, on the compact metric space $M$ has \emph{sensitive dependence} if there exists a $\rho > 0$ such that for each $\delta > 0$ (no matter how small) and each $x \in M$, there is a $y \in M$ that is within the distance $\delta$ of $x$ and for which at some later time $t$, $|f^t(x)-f^t(y)| > \rho$.

That is, no matter how close together the initial conditions $x$ and $y$ are, if we wait long enough, the orbits from these initial conditions will separate by more than some fixed value $\rho$.  Notice that this definition of sensitive dependence does not say anything about the rate at which these orbits diverge from each other: this rate might, for example, be exponential (e.g., as for situations with a positive Lyapunov exponent), or linear (e.g., as for the example in Sec.~\ref{sec2.6.3}).

Another often used concept assigns sensitive dependence to the dynamics on a compact invariant set (the space $M$ is now not necessarily compact) as follows.

\textbf{Definition}:  A continuous map $f$ has sensitive dependence on a compact invariant set $J$ of a metric space $M$ if for every $\delta > 0$ (no matter how small) and every point $x \in J$, there is a point $y \in J$ within a distance $\delta$ of $x$ such that, at some later time $t$, $|f^t(x)-f^t(y)| > \rho$ for some fixed value $\rho > 0$.
 
The following is a definition of chaos, based on that given by Devaney\cite{D}.

\textbf{Definition of Devaney-chaos}:  A continuous map $f:M\to M$, with $M$ a compact metric space, is chaotic if it satisfies the following three conditions.
\begin{enumerate}
\item[(i)] $f$ has sensitive dependence.
\item[(ii)] $f$ has periodic orbits that are dense in $M$.
\item[(iii)] $f$ has an orbit that is dense in $M$, i.e., there exists an initial condition $x^*$ such that for each $y\in M$ and each $\delta > 0$ (no matter how small), at some time $t$, the orbit from $x^*$ will be within the distance $\delta$ from $y$: $|f^t(x^*) - y| < \delta$.
\end{enumerate}

This definition can be converted to define Devaney chaos for a compact invariant set, $J=f(J)$, by replacing $M$ in conditions (ii) and (iii) by $J$, and condition (i) by ``$f$ has sensitive dependence on $J$''.

It was pointed out by Banks et al.~\cite{BBCDS} that conditions (ii) and (iii) of Devaney's definition, imply his condition (i).   Thus condition (i) for Devaney-chaos can be omitted.  A serious drawback of Devaney's definition of chaos is that it excludes significant cases that are sometimes considered and that most would regard as chaotic.  For example, consider a map with quasi-periodic forcing,
\begin{equation}
\label{eq5}
z_{t+1} = G(z_t, \theta_t), \quad \theta_{t+1} = [\theta_t + \omega] \text{ mod } 2\pi,
\end{equation}
where $\omega/(2\pi)$ is an irrational number.  Regarding this as a dynamical system with a state $x = (z,\theta)$, we see that, because of the quasi-periodic behavior of $\theta$, there are no periodic orbits of this system.  Hence, the system (\ref{eq5}) fails condition (ii) for Devaney chaos.  Thus  according to the definition of Devaney chaos, a system like (\ref{eq5}) can never exhibit chaos.  Yet quasi-periodically forced systems are of practical interest and can have attractors with a positive Lyapunov exponent, a situation generally thought of as chaotic.  Another point to make in connection with the example (\ref{eq5}) is that it presents a problem for the Devaney definition of chaos even when $G$ is independent of $\theta$:  in that case $z_{t+1} = G(z_t)$, on its own, might indeed satisfy the conditions for Devaney-chaos; however, by considering the state to be $x=(z,\theta)$ with $\theta_t$ quasi-periodic, the Devaney chaos condition (ii) is not satisfied, even though there is no change in the chaotic dynamics of $z$.

According to Banks et al., Devaney-chaos only requires satisfaction of the two conditions that there be a dense orbit and a dense set of periodic orbits.  Robinson\cite{R}, on the other hand, notes that of the three conditions originally specified by Devaney, the requirement of a dense set of periodic orbits does not seem as ``central to the idea of chaos'' as the other two conditions (sensitive dependence and a dense orbit).  Thus he (and also, independently, Wiggins\cite{W}) proposes the following definition.  

\textbf{Definition of Robinson-chaos}:  The same as Devaney-chaos except that condition (ii) is deleted.

This definition, by not requiring periodic orbits, has the benefit of potentially allowing more consistent treatment of forced systems, like (\ref{eq5}).  However, there is still, in our opinion, a drawback.  This occurs, e.g., with reference to the shear map example (\ref{eqY}) of Sec.~\ref{sec2.6.3},
which was considered by Robinson\cite{R} (see also Hunt and Ott\cite{HO}).  As discussed in Sec.~\ref{sec2.6.3}, orbits are dense and nearby points typically separate linearly with $t$.  Thus, this example is Robinson chaotic.  However, the two Lyapunov exponents are zero.   While this example satisfies the Robinson-chaos definition, due to the slow, linear-in-time, separation of orbits, such dynamics has previously been classified as nonchaotic (see literature on so-called strange nonchaotic attractors\cite{FKP}).  Indeed, this linear-in-time separation of nearby orbits presents comparatively little prediction difficulty as compared to the exponential divergence emphasized by Lorenz. 

Li and Yorke\cite{LY} define the notion of a ``scrambled set'', and the presence of a scrambled set can be taken as another definition of chaos.  While this works well in the original context of one-dimensional maps considered by Li and Yorke, as we will see, it is not as appropriate for higher dimensional systems.

\textbf{Definition of a scrambled set}: For $f:M\to M$ with $M$ a compact metric space, an uncountably infinite subset $J$ of $M$ is scrambled if, for every pair $x,y \in J$ with $x \neq y$,
$$
\limsup_{t\to\infty} |f^t(x) - f^t(y)| > 0, \quad
\liminf_{t\to\infty} |f^t(x) - f^t(y)| = 0.
$$

Thus, by the second Li-Yorke condition, the orbits from $x$ and $y$ come arbitrarily close to each other an infinite number of times, while by the first Li-Yorke condition, the distance between the orbits from $x$ and $y$ also exceeds a fixed positive amount an infinite number of times.  An attractive aspect of scrambling is that it excludes some cases that have sensitive dependence but are usually not considered chaotic.  In particular, the shear map example (\ref{eqY}) discussed above does not have a scrambled set because the $\theta$-distance (or $\phi$-distance, if the $\theta$-distance is $0$) between a pair of orbits remains constant, thus violating the second Li-Yorke condition for scrambling.  Nevertheless, as with Robinson-chaos, the definition of chaos as having an uncountable scrambled set includes cases that are generally regarded as nonchaotic.  One example is a two-dimensional flow with an attracting homoclinic orbit, considered by Robinson\cite{R} and W. Ott and Yorke\cite{OY} (see Figure~1 in either paper); on a trajectory converging to the homoclinic orbit, a finite piece of the trajectory forms an uncountable scrambled set.  Thus, the compact invariant set formed by the homoclinic orbit and its interior exhibits scrambling.

From the discussion above, we see that using notions related to the common definition of sensitive dependence presents problems when attempting to use them to give a generally applicable definition of chaos.  On the other hand, another type of dynamical characterization, namely that of Lyapunov exponents, seems better suited to defining chaos.  Indeed, it can be quite useful to define a chaotic attractor using Lyapunov exponents.  If one excludes certain cases of Milnor attractors (see below) and concentrates on a definition of an attractor of a map $f$ as a bounded set $A$ with a dense orbit such that there is an $\epsilon$-neighborhood $A_\epsilon$ for which $\cap_{t=0}^\infty f^t(A_\epsilon) = A$, then it seems that a good definition of a chaotic attractor of the map $f$ is simply an attractor that has a positive Lyapunov exponent.

\begin{figure}
\includegraphics[width=3.25in]{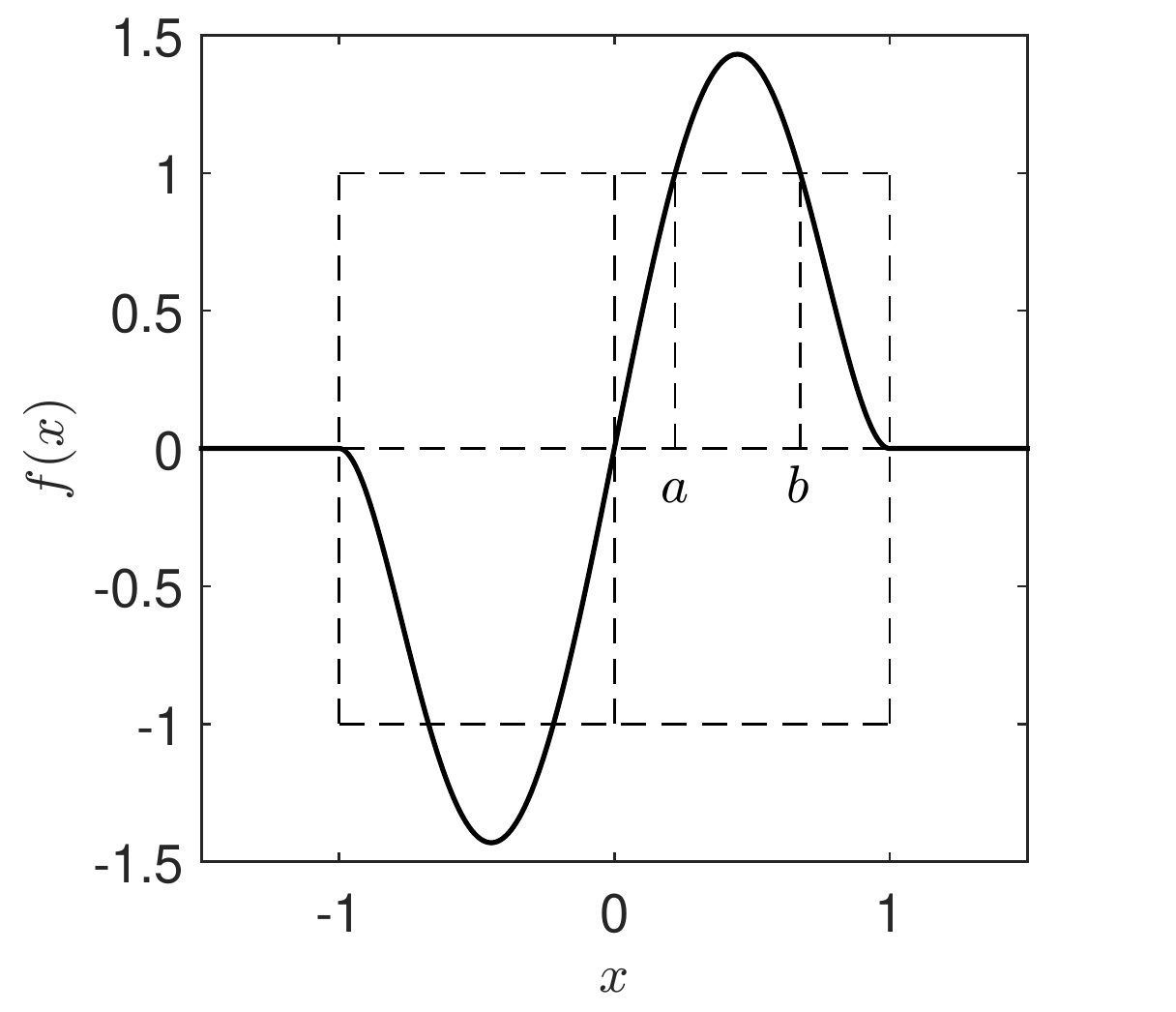}
\caption{\label{fig5}A one-dimensional map for which the unstable fixed point at $x=0$ is a Milnor attractor.  Trajectories that reach $[a,b]$ map to $0$ two iterates later.}
\end{figure}

Now, however, consider Milnor's definition\cite{M} of an attractor: $A$ is an attractor for $f:M\to M$ if there is a positive Lebesgue measure of points $x \in M$ such that $A$ is the forward time limit set of $A$.  Figure~\ref{fig5} shows an example demonstrating that the definition of a chaotic attractor as an attractor with a positive Lyapunov exponent can be problematic, if Milnor's definition of an attractor is used.

The function $f(x)$ in Fig.~\ref{fig5} goes to zero at $x = \pm 1$ and remains zero for $|x|>1$.  There is a positive measure of initial conditions $x_0$ that go to $x=0$ and stay there (e.g., if $a<x_0<b$, then $x_1>0$ and $x_2=x_3=x_4=\cdots =0$).  Thus, the unstable fixed point $x=0$ is a Milnor attractor with a positive Lyapunov exponent (because $df/dx>1$ at $x=0$) yet it would be, we think, unacceptable to call the set $x=0$ chaotic.  This example is rather special and contrived.  Thus, in practice, it is still very useful to think of a chaotic attractor as one with a positive Lyapunov exponent.
However, a main concern of this paper is a definition of chaos that works fairly generally, including being applicable to both attractors and repellers.  In the case of repellers, basing the existence of chaos on Lyapunov exponents presents a problem, since a repelling fixed point with a positive Lyapunov exponent could not reasonably be considered chaotic.
On the other hand, the anomaly for the fixed-point repeller example and the Milnor example of Fig.~\ref{fig5} is removed if we define chaos by positive expansion entropy (see, e.g., Sec.~\ref{sec2.6.1}).

\section{Conclusion}
\label{sec5}

In this paper we have introduced a quantity, the \emph{expansion entropy} (Eqs.~(\ref{eq1}) and (\ref{eq2})), and we have argued that expansion entropy provides a ``good'' definition of chaos in that it possesses several desirable properties.  We also compare this definition with other past definitions of chaos (Secs.~\ref{sec3} and \ref{sec4}).  In particular, the expansion entropy $H_0$ enjoys the properties of \emph{generality}, \emph{simplicity}, and \emph{computability} discussed in Sec.~\ref{sec1}.  One important feature of $H_0$ is that it assesses the presence of chaos in any given bounded region $S$ in state space, rather than in an invariant set.  As such, it applies naturally in cases where the invariant sets are unknown or (e.g., in both deterministically and randomly forced systems) do not exist.  Section~\ref{sec2.6.2} presents examples illustrating various issues and features of expansion entropy, perhaps most importantly its numerical computation.  It is our hope that our paper will lead to the use of expansion entropy in applications and to further study of its properties.

\begin{acknowledgments}
The work of E. Ott was supported by the U.S. Army Research Office under Grant W911NF-12-1-0101.  We thank S. Newhouse for pointing out earlier work related to expansion entropy, and J. Yorke and the reviewers for helpful comments.
\end{acknowledgments}

\appendix

\section{$q$-Order Expansion Entropy}
\label{appA}

As discussed in Sec.~\ref{sec2.5}, we here generalize our definition of $H_0$ to a definition of a $q$-order entropy $H_q$:
\begin{equation}
\label{eq3}
H_q = \frac{1}{1-q} \lim_{t'\to\infty} \frac{1}{t'-t} \ln \left\{\frac{\int_{S_{t',t}} G^{1-q} d\mu}{\mu(S_{t',t})^q \mu(S)^{1-q}}\right\},
\end{equation}
where the argument of $G$ is the same as in Eq.~(\ref{eq1}).  Comparing Eqs.~(\ref{eq1}) and (\ref{eq2}) with (\ref{eq3}), we see that (\ref{eq3}) reduces to (\ref{eq1}) and (\ref{eq2}) for $q=0$.  Furthermore, letting $q\to 1$ and assuming that the $q\to 1$ and $t\to\infty$ limits can be interchanged, we obtain
\begin{equation}
\label{eq4}
H_1 = \left[\lim_{t'\to\infty} \frac{1}{t'-t} \frac{\int_{S_{t',t}} \ln G d\mu}{\mu(S_{t',t})}\right] - \frac{1}{\tau_+},
\end{equation}
where, as in Sec.~\ref{sec2.3},
$$
\frac{1}{\tau_+} = \lim_{t'\to\infty} \frac{1}{t'-t} \ln \left(\frac{\mu(S)}{\mu(S_{t',t})}\right).
$$
The quantity (\ref{eq4}) can be viewed as bearing a relationship to metric entropy that is analogous to the relationship between $H_0$ and topological entropy.
In the case where $S$ contains an attractor, $1/\tau_+ = 0$.  In the case where $S$ contains a repeller, we call $\tau_+$ the average lifetime of repeller orbits.   In the case where $S$ is a neighborhood if an invariant set with a ``natural measure''\cite{HOY},  we can identify the first term in Eq.~(\ref{eq4}) with the sum of the positive Lyapunov exponents $\lambda_j$:
\begin{equation}\label{eqQ}
H_1 = \sum_{\lambda_j > 0} \lambda_j - 1/\tau_+,
\end{equation}
which agrees with the results for metric entropy of Kantz and Grassberger\cite{KG,HOY} for chaotic repellers and of Pesin\cite{P} for $1/\tau_+ = 0$.

We now obtain $H_q(f,S)$ and $H_q(f,S')$ for the example in Sec.~\ref{sec2.6.6}, where $S$ is the large interval $[-1,1.5]$ containing both the attracting fixed point and the chaotic repeller, while $S' \subset S$ is the smaller restraining region $[0,1]$ containing only the chaotic repeller.

\begin{figure}
\includegraphics[width=3.25in]{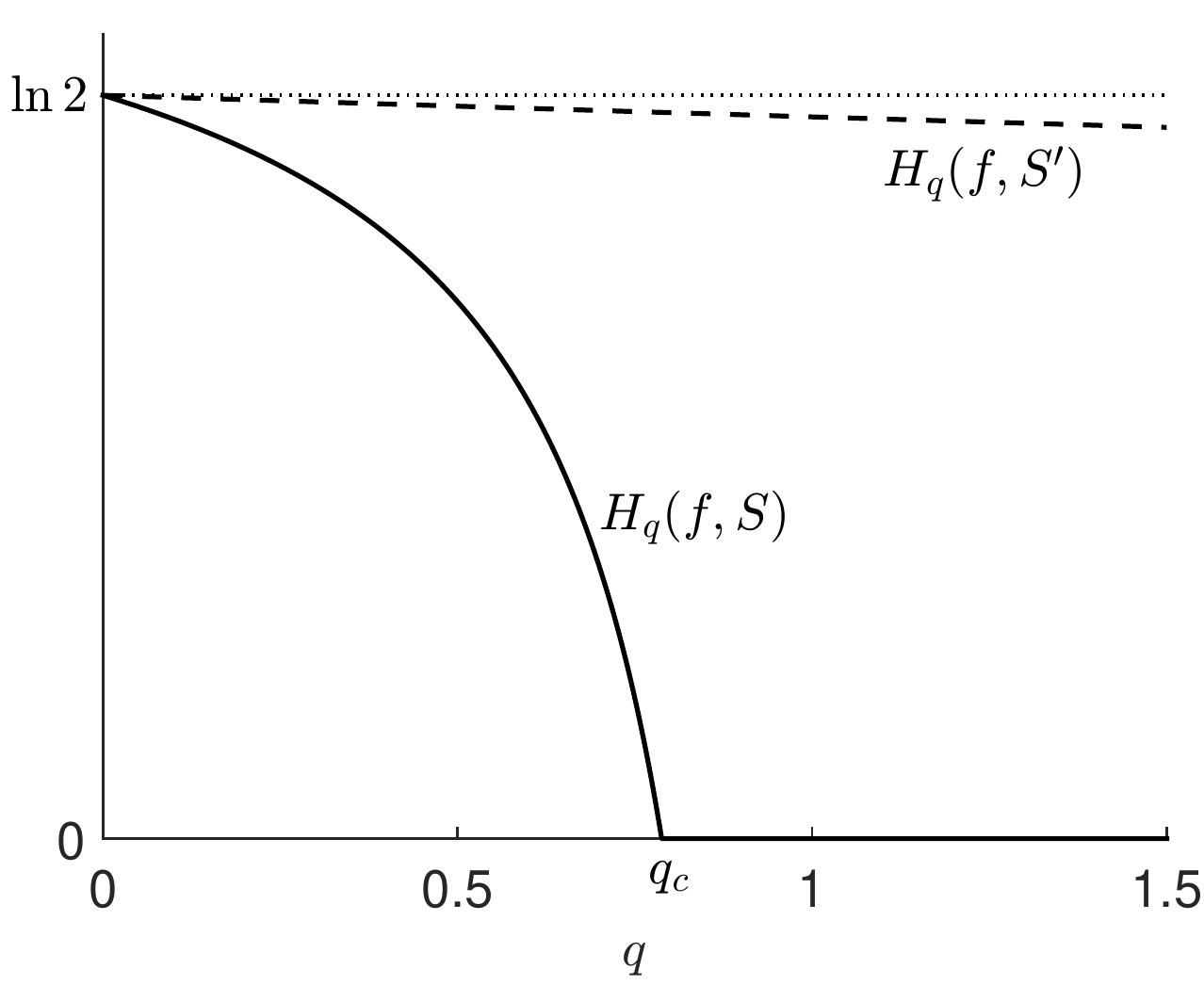}
\caption{\label{fighq}$H_q(f,S')$ (dashed curve) and $H_q(f,S)$ (solid curve) versus $q$.  The dashed curve decreases slightly from $\ln 2 \approx 0.693$ at $q=0$ to about $0.663$ at $q=1.5$.}
\end{figure}

We begin by finding $H_q(f,S')$.  As discussed in Sec.~\ref{sec2.6.6}, the set $S'_{\dt,0}$ of initial conditions that remain in $S'$ from time $0$ to time $T$ consists of $2^\dt$ initial intervals of varying widths $3^{-k} 2^{k-\dt}$ (where $k = 0, 1, \ldots, \dt$) on each of which $G(Df^\dt) = 3^k 2^{\dt - k}$.  In addition, the number of such intervals with a given $k$ is the binomial coefficient $C(\dt,k) = \dt!/[k! (\dt-k)!]$.  Thus, the integral of $G^{1-q}$ that appears in Eq.~(\ref{eq3}) is
\begin{eqnarray}\label{eqB}
\int_{S'_{\dt,0}} G(Df^\dt)^{1-q} d\mu
&=& \sum_{k=0}^\dt C(\dt,k) [3^k 2^{\dt - k}]^{1-q} 3^{-k} 2^{k - \dt} \nonumber\\
&=& (3^{-q} + 2^{-q})^\dt.
\end{eqnarray}
Furthermore, the total length of $S'_{\dt,0}$ decreases by the ratio $5/6$ upon increase of $\dt$ by one, so that
\begin{equation}\label{eqC}
\mu(S'_{\dt,0}) = (5/6)^\dt.
\end{equation}
Using Eqs.~(\ref{eqB}) and (\ref{eqC}) in Eq.~(\ref{eq1}), we obtain
\begin{equation}\label{eqD}
H_q(f,S') = (1-q)^{-1} \ln [(2/5)^q + (3/5)^q].
\end{equation}
Note that this quantity is positive for all $q \geq 0$, and decreases monotonically with increasing $q$ (dashed curve in Fig.~\ref{fighq}).  For $q=0$, this agrees with our previous result of Sec.~\ref{sec2.6.6} that $H_0(f,S')=\ln 2$, while taking the limit $q\to 1$ yields $H_1(f,S') = (2/5)\ln(5/2) + (3/5)\ln(5/3) > 0$.  As $q\to\infty$, $H_q(f,S')\to \ln(5/3)$, so that $H_0 = \ln 2 \geq H_q \geq H_\infty = \ln(5/3)$.

We now turn to the evaluation of the $q$-order expansion entropy for the larger restraining region $S = [-1,1.5]$.  The main difference from $S'$ is that $S_{\dt,0} = S$ for all $\dt \geq 0$, so $\mu(S_{\dt,0}) = 2.5$, in contrast with Eq.~(\ref{eqC}).  To estimate the integral of $G(Df^\dt)^{1-q}$ over $S_{\dt,0}$, note first that by Eq.~(\ref{eqB}),
\begin{eqnarray*}
\int_{S_{\dt,0}} G(Df^\dt)^{1-q} d\mu
&\geq& \int_{S'_{\dt,0}} G(Df^\dt)^{1-q} d\mu \\
&=& (3^{-q} + 2^{-q})^\dt.
\end{eqnarray*}
The contribution to the integral of $G(Df^\dt)^{1-q}$ from initial conditions in $S_{\dt,0}$ but not in $S'_{\dt,0}$ can be bounded above by $C\dt\max[(3^{-q} + 2^{-q})^\dt,1]$ for a constant $C$ independent of $\dt$; the factor of $\dt$ in the upper bound comes from considering trajectories that first leave $S'$ at time $t$, for each of the values $t = 0, 1, \ldots, \dt-1$.  Also, since $G(Df^\dt) = 1$ for initial conditions in the interval near $x = -1/2$ on which $Df < 1$, such initial conditions contribute at least $c > 0$ to the integral, where $c$ is the length of the contracting interval.	 Thus,
\begin{eqnarray*}
c + (3^{-q} + 2^{-q})^\dt
&\leq& \int_{S_{\dt,0}} G(Df^\dt)^{1-q} d\mu \\
&\leq& C\dt \max[(3^{-q} + 2^{-q})^\dt,1].
\end{eqnarray*}
From Eq.~(\ref{eq3}), recalling that $\mu(S_{\dt,0}) = \mu(S)$, we conclude that for $q \neq 1$,
\begin{equation}\label{eqH}
H_q(f,S) = (1-q)^{-1} \ln \max(3^{-q}+2^{-q},1).
\end{equation}
Note that there is a critical value $0 < q_c < 1$ for which $3^{-q_c} + 2^{-q_c} = 1$; then $H_q(f,S) = 0$ for $q \geq q_c$.  In particular, $H_1(f,S) = 0$ by taking the limit $q\to 1$.

Comparing Eqs.~(\ref{eqD}) and (\ref{eqH}), we see that $H_0(f,S') = H_0(f,S) = \ln 2$ (as we argued in Sec.~\ref{sec2.6.2}), but $H_q(f,S') > H_q(f,S)$ for $q > 0$; see Fig.~\ref{fighq}.  Note also that if the slopes $3$ and $2$ were increased, the critical value $q_c$ beyond which $H_q(f,S) = 0$ could be made arbitrarily close to $0$.  We conclude that $H_q$ for $q > 0$ does not always detect chaos (i.e., $H_q$ may be zero) in a restraining region containing an invariant set that is chaotic by all the definitions we reviewed in Sec.~\ref{sec4}, as well as by our definition $H_0 > 0$.

\section{Relation of Expansion Entropy Integral to Topological Entropy Ratio}
\label{appB}

Here we justify the claim in Section~\ref{sec3} that the integral $E_{T,0}(f,S)$ of Eq.~(\ref{eq1}) used to define expansion entropy approximates, for small $\epsilon$, the ratio $\tilde{N}(T,\epsilon)/N(0,\epsilon)$ related to the definition of topological entropy.  Below, when we say that two quantities have the ``same order of magnitude'', we mean that their ratio is bounded above and below by positive constants that are independent of $\epsilon$ and $T$.

Assume that $\epsilon > 0$ is small enough that the remainder term in the first order Taylor expansion of $f_{T,0}$ is much smaller than $\epsilon$ for points within $\epsilon$ of each other:
$$
|f_{T,0}(y) - f_{T,0}(x) - Df_{T,0}(x)(y-x)| \ll \epsilon
$$
for $x,y \in S$ with $|y-x| \leq \epsilon$.  Cover $S$ with a grid of $N_0$ boxes whose diameters are $\epsilon$; then $N_0$ has the same order of magnitude as the maximum number $N(0,\epsilon)$ of $\epsilon$-separated points in $S$.  Each box $B$ is contained in a ball of radius $\epsilon$, and contains a ball whose radius has the same order of magnitude as $\epsilon$.  Notice also that $\mu(B) \approx \mu(S)/N_0$ for small $\epsilon$.  Let $x_B$ be the center of $B$, and let $\sigma_1 \geq \sigma_2 \geq \cdots \geq \sigma_n$ be the singular values of $Df_{T,0}(x_B)$.  Then the image of $B$ under $f_{T,0}$ contains and is contained in ellipses whose semiaxes have the same order of magnitude as $\sigma_1 \epsilon, \sigma_2 \epsilon, \ldots, \sigma_n \epsilon$.  Let $d$ be the largest index for which $\sigma_d > 1$.  Then the maximum number of $\epsilon$-separated points in the image of $B$ has the same order of magnitude as $\sigma_1 \sigma_2 \cdots \sigma_d = G(Df_{T,0}(x_B))$.  Summing over all $B$, the maximum number $\tilde{N}(T,\epsilon)$ of trajectories that are $\epsilon$-separated at either time $0$ or at time $T$ has the same order of magnitude as
\begin{eqnarray*}
\sum_B G(Df_{T,0}(x_B))
&\approx& \frac{1}{\mu(B)} \int_S G(Df_{T,0}(x)) d\mu(x) \\ &\approx& \frac{N_0}{\mu(S)} \int_S G(Df_{T,0}(x)) d\mu(x)
\end{eqnarray*}

Since we assumed that $S$ is invariant, $S_{T,0}$ is the same as $S$.  Comparing with Eq.~(\ref{eq1}), the discussion above constitutes an outline of a proof that $\tilde{N}(T,\epsilon)/N(0,\epsilon)$ has the same order of magnitude as $E_{T,0}(f,S)$.  (In fact, the same is true when $S$ is not invariant, if we define $\tilde{N}$ to count only trajectories that remain in $S$ between times $0$ and $T$.)


\begin{thebibliography}{99}

\bibitem{LY}
T. Y. Li and J. A. Yorke,
Period three implies chaos,
Amer. Math. Monthly 85, 985--992 (1975).

\bibitem{L}
E. Lorenz,
Deterministic nonperiodic flow,
J. Atmos. Sci. 20, 130--141 (1963).

\bibitem{GOY}
C. Grebogi, E. Ott and J. A. Yorke,
Crises, sudden changes in chaotic attractors and chaotic transients,
Physica D 7, 181--200 (1983).

\bibitem{KG}
H. Kantz and P. Grassberger,
Repellers, semi-attractors and long-lived chaotic transients,
Physica D 17, 75--86 (1985).

\bibitem{LT}
Y.-C. Lai and T. T\'el,
\emph{Transient Chaos: Complex Dynamics on Finite Time Scales},
Applied Mathematical Sciences 173 (Springer, New York, 2011).

\bibitem{MGOY}
S. W. McDonald, C. Grebogi, E. Ott, and J. A. Yorke,
Fractal basin boundaries,
Physica D 17, 125--153 (1985).

\bibitem{TO}
For a review see T. Tel and E. Ott,
Chaotic scattering: An introduction,
Chaos 3, 417--426 (1993).

\bibitem{SYE}
J. D. Skufca, J. A. Yorke, and B. Eckhardt,
Edge of chaos in parallel shear flow,
Phys. Rev. Lett. 96, 174101 (2006).

\bibitem{SKG}
J. C. Sommerer, H. C. Ku and H. E. Gilrath,
Experimental evidence for chaotic scattering in a fluid wake,
Phys. Rev. Lett. 77, 5055--5058 (1996).

\bibitem{JRLMFU}
C. Jaffe, S.D. Ross, M. L. Lo, J.Marsden, D. Farrelly and T. Uzer,
Statistical theory of asteroid escape rates,
Phys. Rev. Lett. 89, 011101 (2002).

\bibitem{GE}
e.g., R. E. Gillian and G. S. Ezra,
Transport and turnstiles in multidimensional Hamiltonian mappings for unimolecular fragmentation: Application to van der Waals predissociation, J. Chem. Phys. 94, 2648--2668 (1991).

\bibitem{DD}
e.g., M. L. Du and J. Delos,
Effect of closed classical orbits on quantum spectra: ionization of atoms in a magnetic field I: Physical picture and calculations,
Phys. Rev. A 38, 1896--1912 (1988).

\bibitem{FKP}
U. Feudel, S. Kuznetsov and A. Pikovsky,
\emph{Strange Nonchaotic Attractors} (World Scientific, Singapore, 2006).

\bibitem{LY1}
F. Ledrappier and L.-S. Young,
Dimension formula for random transformations,
Comm. Math. Phys. 117, 529--548 (1988).

\bibitem{LY2}
F. Ledrappier and L.-S. Young,
Entropy formula for random transformations,
Prob. Th. Rel. Fields 80, 217--240 (1988).

\bibitem{YOC}
L. Yu, E. Ott, Q. Chen,
Transition to chaos for random dynamical systems,
Phys. Rev. Lett. 65, 2935--2938 (1990).

\bibitem{RBKOU}
I. I. Rypina, F. J. Beron-Vera, M. G. Brown, H. Kocak, M. J. Olascoaga and I. A. Udovydchenkov,
On the Lagrangian dynamics of atmospheric zonal jets and the permeability of the stratospheric polar vortex,
J. Atmos. Sci. 64, 3595--3610 (2007).

\bibitem{LKKHLD}
J. F. Lindner, V. Kohar, B. Kia, M. Hippke, J. G. Learned and W. L. Ditto, Strange non-chaotic stars, Phys. Rev. Lett. 113, 054101 (2015).

\bibitem{Mo}
P. Moskalik, Multimode oscillations in classical Cepheids and RR Lyrae-type stars, Proc. Int. Astron. Union 9(S301), 249--256. (2013).

\bibitem{VAO}
F. Varosi, T. M. Antonsen and E. Ott,
The spectrum of fractal dimensions of passively convected scalar gradients in chaotic fluid flows, Phys. Fluids A 257, 1017--1028 (1993).

\bibitem{SO}
J. C. Sommerer and E. Ott,
Particles foating on a moving fluid: a dynamically comprehensible phusical fractal,
Sci. 257, 335--339 (2005).

\bibitem{HY}
G. Haller and G. Yuan,
Lagrangian coherent structures in three-dimensional fluid flows,
Phys. D 149, 352--370 (2000).

\bibitem{VHG}
G. A. Voth, G. Haller and J.P. Gollub,
Experimental measurements of stretching fields in fluid mixing,
Phys. Rev. Lett. 88, 254501 (2002).

\bibitem{MHPRS}
M. Mathur, G. Haller, T. Peacock, J. E. Ruppert-Felsot and H. L. Swinney,
Uncovering the Lagrangian skeleton of turbulence,
Phys. Rev. Lett. 98, 144501 (2007).

\bibitem{PROK}
e.g., A. S. Pikovsky, M. G. Rosenblum, G. V. Osipov and J. Kurths,
Phase synchronization of chaotic oscillators by external driving,
Physica D 104, 219--238 (1997).

\bibitem{SS}
R. Sacksteder and S. Shub,
Entropy of a differentiable map,
Adv. Math. 28, 181--185 (1978).

\bibitem{K}
O. S. Kozlovski,
An integral formula for topological entropy of $C^\infty$ maps,
Erg. Th. Dynam. Sys. 18, 405--424 (1998).

\bibitem{footnote1}
If $M$ is a Riemannian manifold, then it has a canonical volume that is equivalent to Lebesgue measure in appropriate local coordinates.  Furthermore, when we treat the derivative of a map as a matrix, or use (small) distances in $M$, we assume the use of ``normal coordinates'', which exist at least for $C^2$ Riemannian manifolds.

\bibitem{footnote2}
Though we define expansion entropy only for a particular realization of a stochastic system, we expect that under appropriate hypotheses it has the same value for almost every realization.
At a minimum, this should be the case when $S$ is invariant for all realizations of a system forced by an IID process, because in this case the expansion entropy depends only on the tail of the IID process.
A suitable setting for the rigorous study of expansion entropy in random systems would be that of Ledrappier and Young \cite{LY1,LY2}.

\bibitem{footnote3}
If $f$ is a $C^\infty$ map on a compact manifold, and $S$ is the entire manifold, then the limit has been proved to exist\cite{K}.  More generally, $H_0$ could be defined as the $\limsup$, as in other definitions of entropy (see Sec.~\ref{sec3}).  

\bibitem{JOH}
J. Jacobs, E. Ott, B. R. Hunt,
Calculating topological entropy for transient chaos with an application to communicating with chaos,
Phys. Rev. E 57, 6577-6588 (1998).

\bibitem{BR}
J. Balatoni and A. Renyi, Remarks on entropy, Pub. Math. Inst. Hung. Acad. Sci. 1, 9 (1956).  Translated in \emph{Selected Papers of A. Renyi}, vol. 1, p. 558 (Academiai, Budapest, 1976).

\bibitem{HP}
H. G. E. Hentschel and I. Procaccia,
The infinite number of generalized dimensions of fractals and strange attractors,
Physica D 8, 435--444 (1983).

\bibitem{GP2}
P. Grassberger and I. Procaccia,
Dimensions and entropies of strange attractors from a fluctuating dynamics approach,
Physica D 13, 34--54 (1984).

\bibitem{GP1}
P. Grassberger and I. Procaccia,
Estimation of Kolmogorov entropy from a chaotic signal,
Phys. Rev. A 28, 2591--2593 (1983).

\bibitem{HOY}
B. R. Hunt, E. Ott and J. A. Yorke,
Fractal dimensions of chaotic saddles of dynamical systems,
Phys. Rev. E 54, 4819--4823 (1996).

\bibitem{P}
Ya. B. Pesin,
Lyapunov characteristic exponents and ergodic properties of smooth dynamical systems with an invariant measure, Dokl. Akad. Nauk SSSR 226, 774--777 (1976).  Translated in 
Sov. Math. Dokl. 17, 196--199 (1976).

\bibitem{R}
C. Robinson, What is a Chaotic Attractor?, Qualitative Theory of Dynamical Systems 7, 227--236 (2008).

\bibitem{HO}
B. R. Hunt and E. Ott, Fractal Properties of Robust Strange Non-Chaotic Attractors, Phys. Rev. Lett. 87, 254101 (2001).

\bibitem{DN}
R. L. Devaney and Z. Nitecki, Shift automorphisms in the H\'enon mapping, Comm. Math. Phys. 67, 137--146 (1979).

\bibitem{AKM}
R. L. Adler, A. G. Konheim and M. H. McAndrew,  Topological entropy, Trans. Amer. Math. Soc. 114, 309--319 (1965).

\bibitem{Din}
E. I. Dinaburg, The relation between topological entropy and metric entropy, Dokl. Akad. Nauk SSSR 190, 19--22 (1970) [translated from Soviet Math. Dokl. 11, 13--16 (1969)].

\bibitem{Bo1}
R. Bowen: Entropy for group endomorphisms and homogeneous spaces, Trans. Amer. Math. Soc. 153, 401--414 (1971).

\bibitem{Bo2}
R. Bowen, Periodic points and measures for axiom A diffeomorphims, Trans. Amer. Math. Soc. 154, 377--397 (1971).

\bibitem{Pr}
F. Przytycki,
An upper estimation for topological entropy of diffeomorphisms,
Inv. Math. 59, 205--213 (1980).

\bibitem{N}
S. Newhouse,
Entropy and Volume,
Erg. Th. Dynam. Sys. 8, 283--299 (1988).

\bibitem{MS}
M. Misiurewicz and W. Szlenk,
Entropy of piecewise monotone mappings,
Stud. Math. 57, 45--63 (1980).

\bibitem{Y}
Y. Yomdin,
Volume Growth and Entropy,
Israel J. Math. 57, 285--300 (1987).

\bibitem{G}
M. Gromov,
Entropy, homology and semialgebraic geometry, S\'em. Bourbaki 28, 225--240 (1985-6).

\bibitem{NP}
S. Newhouse and T. Pignataro,
On the Estimation of Topological Entropy,
J. Stat. Phys. 72, 1331--1351 (1993).

\bibitem{KT}
Z. Kov\'acs and T. T\'el, 
Thermodynamics of irregular scattering,
Phys. Rev. Lett. 64, 1617--1620 (1990).

\bibitem{COH}
Q. Chen, E. Ott, and L. P. Hurd,
Calculating topological entropies of chaotic dynamical systems,
Phys. Lett. A 156, 48--52 (1991).

\bibitem{FJO}
G. Froyland, O. Junge, and G. Ochs,
Rigorous computation of topological entropy with
respect to a finite partition,
Physica D 154, 68-–84 (2001); see in particular Remark B.2.

\bibitem{D}
R. L. Devaney, \emph{An Introduction to Chaotic Dynamical Systems}, (Addison-Wesley, New York and Reading 1989).

\bibitem{BBCDS}
J. Banks, J. Brooks, G. Cairns, G. Davis, and P. Stacy, On Devaney's Definition of Chaos, Amer. Math. Monthly 99, 332--334 (1992).

\bibitem{W}
S. Wiggins, \emph{Chaotic Transport in Dynamical Systems}, Interdisciplinary Applied Mathematics Series, Volume 2 (Springer-Verlag, Berlin, 1992).

\bibitem{OY}
W. Ott \& J. Yorke,
When Lyapunov exponents fail to exist,
Phys. Rev. E 78, 056203 (2008).

\bibitem{M}
J. Milnor, On the Concept of an Attractor, Comm. Math. Phys. 99, 177--195 (1985).

\end{thebibliography}

\end{document}